\documentclass[twocolumn,aps,pra,amsmath,superscriptaddress,notitlepage,longbibliography]{revtex4-1}
\usepackage{amssymb}
\usepackage{mathrsfs}
\usepackage{graphicx}
\usepackage{float}
\usepackage[caption=false]{subfig}
\usepackage{epstopdf}

\usepackage[normalem]{ulem}
\usepackage{verbatim}
\usepackage{xcolor}
\usepackage{bm}
\usepackage[utf8x]{inputenc}
\usepackage[colorlinks,citecolor=blue]{hyperref}

\begin{document}
%\linenumbers

\title{Light-enhanced nonlinear Hall effect}
\author{Fang Qin}
\email{qinfang@nus.edu.sg}
\affiliation{Department of Physics, National University of Singapore, Singapore 117551, Singapore}
\author{Rui Chen}
\email{chenr@hubu.edu.cn}
\affiliation{Department of Physics, Hubei University, Wuhan 430062, China}
\author{Ching Hua Lee}
\email{phylch@nus.edu.sg}
\affiliation{Department of Physics, National University of Singapore, Singapore 117551, Singapore}

\date{\today}

\begin{abstract}
%\begin{linenumbers}
It is well known that a nontrivial Chern number results in quantized Hall conductance. What is less known is that, generically, the Hall response can be dramatically different from its quantized value in materials with broken inversion symmetry. This stems from the leading Hall contribution beyond the linear order, known as the Berry curvature dipole (BCD). While the BCD is in principle always present, it is typically very small outside of a narrow window close to a topological transition and is thus experimentally elusive without careful tuning of external fields, temperature, or impurities. In this work, we transcend this challenge by devising optical driving and quench protocols that enable practical and direct access to large BCD and nonlinear Hall responses. Varying the amplitude of an incident circularly polarized laser drives a topological transition between normal and Chern insulator phases, and importantly allows the precise unlocking of nonlinear Hall currents comparable to or larger than the linear Hall contributions. This strong BCD engineering is even more versatile with our two-parameter quench protocol, as demonstrated in our experimental proposal. Our predictions are expected to hold qualitatively across a broad range of Hall materials, thereby paving the way for the controlled engineering of nonlinear electronic properties in diverse media.
%\end{linenumbers}
\end{abstract}
\maketitle

%%%%%%%%%%%%%%%%%%%%%%%%%%%%%%%%%%%%%%%%%%%%%%%%%%%%%%%%%%%%%%%%%%%%%%%%%%%%%%%%%%%%%%%%%%%%%%%%%%%%%%%%%%%%%%
%\onecolumngrid

\section{Introduction}\label{1}

The nonlinear Hall effect is an exciting response contribution beyond the much-studied quantum anomalous Hall effect~\cite{chang2013experimental,sodemann2015quantum,du2018band,du2019disorder,du2021nonlinear,ortix2021nonlinear,bandyopadhyay2024non,ma2019observation}, in which an applied electric field results in a quantized transverse current response in an insulator. It stems from a higher-order correction in the quantum response known as the Berry curvature dipole (BCD)~\cite{sodemann2015quantum}, which generically exists whenever inversion symmetry is broken~\cite{sodemann2015quantum,du2018band,du2019disorder,du2021nonlinear}. Venturing beyond the topological paradigm, the BCD leads to an interesting nonlinear anisotropic current response accompanied by higher harmonic generation~\cite{mrudul2021high,bharti2022high,bharti2023massless,lv2021high,ghimire2019high,nourbakhsh2021high,lee2015negative,tai2021anisotropic} and has motivated a large number of experimental investigations. Indeed, the BCD has been observed in a variety of materials, such as two-dimensional bilayer or few-layer WTe$_2$~\cite{ma2019observation,xu2018electrically,kang2019nonlinear,xiao2020berry,ye2023control,huang2023giant}, BiTeI~\cite{facio2018strongly}, bilayer graphene~\cite{ho2021hall}, twisted bilayer graphene~\cite{duan2022giant,huang2023intrinsic}, twisted double bilayer graphene~\cite{sinha2022berry}, strained twisted bilayer graphene~\cite{pantaleon2021tunable,zhang2022giant}, two-dimensional monolayer WSe$_2$~\cite{qin2021strain}, strained twisted bilayer WSe$_2$~\cite{hu2022nonlinear}, two-dimensional monolayer MoS$_2$~\cite{lee2017valley,son2019strain}, Weyl semimetal TaIrTe$_4$~\cite{kumar2021room}, three-dimensional Dirac semimetal Cd$_3$As$_2$~\cite{shvetsov2019nonlinear,zhao2023gate}, $\mathbb{Z}_2$ topological insulator Bi$_2$Se$_3$~\cite{he2021quantum}, the Weyl--Kondo semimetal Ce$_3$Bi$_4$Pd$_3$~\cite{dzsaber2021giant}, and T$_{d}$-MoTe$_2$~\cite{tiwari2021giant}. The dominance of the intrinsic nonlinear Hall effect in antiferromagnets with ${\cal P}{\cal T}$ symmetry has been studied~\cite{ma2023anomalous,das2023intrinsic,liu2021intrinsic,gao2014field,wang2021intrinsic} and subsequently experimentally probed in thick T$_{d}$-MoTe$_2$ samples~\cite{lai2021third}. More recently, the quantum metric-induced nonlinear Hall effect in a topological antiferromagnet MnBi$_2$Te$_4$ has also been reported~\cite{gao2023quantum,kaplan2024unification,wang2023quantum}.

Yet, despite its supposed ubiquity, observing the BCD (or its associated nonlinear Hall response) is fraught with practical challenges. Besides generic experimental demands such as maintaining low temperatures ($10\sim100$ K~\cite{ma2019observation,du2021nonlinear}), a key challenge is that the BCD is only significant in a narrow window very close to a topological phase transition~\cite{du2018band}. Accessing it hence requires precise control of external parameters such as magnetic field~\cite{gao2023quantum,wang2023quantum}, electric field~\cite{xu2018electrically,xiao2020berry,ma2019observation,du2021nonlinear,lai2021third}, temperature~\cite{ma2019observation,lai2021third,gao2023quantum,wang2023quantum}, charge/carrier density~\cite{ma2019observation,gao2023quantum,wang2023quantum}, strain~\cite{pantaleon2021tunable,zhang2022giant,sinha2022berry,qin2021strain,hu2022nonlinear}, or impurities~\cite{du2019disorder,duan2022giant,huang2023intrinsic,chen2024nonlinear,atencia2023disorder}, which are not easily tuned. As such, only a few notable experimental observations of the nonlinear Hall effect exist till date, through measuring the BCD~\cite{ma2019observation,xu2018electrically,kang2019nonlinear,xiao2020berry,ye2023control,huang2023giant,ho2021hall,sinha2022berry}, intrinsic nonlinear Hall conductivity~\cite{lai2021third,gao2023quantum,wang2023quantum,huang2023intrinsic}, or both~\cite{duan2022giant}.

To circumvent these challenges, we devise optical driving and quench protocols that can precisely tune a system such that the nonlinear Hall response is dramatically enhanced. Through the periodic driving of polarized light, we not only obtain an approach for driving topological phase transitions that may be inaccessible in static systems, but more importantly unlock new routes towards the versatile engineering of the BCD. Obtaining the effective Hamiltonian through the high-frequency Floquet-Magnus expansion~\cite{magnus1954exponential,blanes2009magnus,lee2018floquet}, we identify a light-enhanced peak in the BCD that can be tuned via the intensity of an applied circularly polarized laser in a very versatile manner. This is accompanied by a critical light-induced band inversion, which demarcates a topological phase boundary between Chern and normal insulator phases, notably without involving the tuning of any other physical parameter. Crucially, the precise tunability of our approach allows for further band engineering via temporal modulation of the laser strength, resulting in controlled modulation of the BCD strength over a wide range.

\section{Results}\label{2}
\subsection{General formalism}

We first provide the foundational framework for the nonlinear Hall effect, such as to contrast it with the more commonly known linear Hall effect. When subjected to external driving forces, the typical dominant response observed is the linear response, which, for instance, accounts for the well-known quantum Hall effect with broken time-reversal symmetry. In general, a linear response encompasses longitudinal and transverse currents. However, nonlinear response can also manifest strongly in transverse currents~\cite{sodemann2015quantum,kang2019nonlinear,ma2019observation,gao2014field,wang2021intrinsic,du2018band,du2019disorder,du2021nonlinear,chen2024nonlinear,atencia2023disorder,lai2021third,huang2023intrinsic,gao2023quantum,wang2023quantum} due to higher-order Berry curvature effects, with a doubled frequency component in the current signal in the transverse direction. Consequently, the transverse current is proportional to the second power in the longitudinal fields.

The response from a driven electric field $\bold E$ is measurable through the electric current density $\bold J$. We expand the electric current density $J_i$ in increasing orders of the electric field as
\begin{equation}\label{eq:v}
J_{i}=\sigma_{ij}E_{j}+\chi_{ijk}E_{j}E_{k}+\xi_{ijkl}E_{j}E_{k}E_{l}+\cdots,
\end{equation} where $\{i,j,k,l\}\in\{x,y,z\}$, and $\mathbf{E}=(E_{x},E_{y},E_{z})$ is the external electric field. Here, the first term denotes the linear response, and its frequency follows the external electric field. The second term is the leading nonlinear contribution with a doubled frequency.

Below, we shall derive the linear and nonlinear responses through a semi-classical approach, where the perturbative effects of an external electric field are treated via the electronic occupation functions.
Subsuming the effects of electron scattering under a phenomenological relaxation time $\tau$, the relaxation time approximation yields the following Boltzmann equation~\cite{schliemann2003anisotropic,sinitsyn2007semiclassical,sodemann2015quantum,mahan2013many}
\begin{equation}
-\tau\partial_{t}f + \frac{e\tau}{\hbar}E_{a}\partial_{k_a}f=f-f_{0},
\end{equation}
where $-e$ is the electron charge, $\hbar$ is the reduced Planck\rq{}s constant, $k_{a}$ is the wave vector of electronic wavepackets along the $a\in\{x,y,z\}$ directions, $f_{0}$ is the equilibrium occupancy given by the Fermi-Dirac distribution function, and $f$ is the non-equilibrium distribution function. Here, $\tau$ sets the timescale for $f$ to relax to its equilibrium distribution $f_0$. By inverting the derivative operators, we can directly solve for the non-equilibrium distribution function $f$ as follows:
\begin{equation}
f\!=\!\frac{f_0}{1\!+\!\tau\partial_{t}\!-\!\frac{e\tau}{\hbar}E_{a}\partial_{k_a}}
\!\approx\!\sum_{n=0}^{\infty}\left(\!-\tau\partial_{t}\!+\!\frac{e\tau}{\hbar}E_{a}\partial_{k_a} \right)^{n}f_{0}.
\end{equation}
To understand how the response originates from this mathematical framework, we expand $f$ into $f={\rm Re}(f_{0}+f_{1}+f_{2}+\cdots)$, where $f_j$ contains terms with the $j$-th power of the electric fields.  Below, we shall only retain up to the quadratic order, such that
\begin{equation}
{\bf J}({\bf E})
\!\approx\!-e\int\frac{d^{2}{\bf k}}{(2\pi)^2}\left(\!\frac{1}{\hbar}\nabla_{\bf k}\epsilon_{\bf k}^{} \!+\! \frac{e}{\hbar}{\bf E}\!\times\!{\bf\Omega}_{\bf k}\!\right){\rm Re}(f_{0}\!+\!f_{1}\!+\!f_{2}),
\label{eq:JE}
\end{equation}
with band dispersion $\epsilon_{\bf k}^{}$ and the Berry curvature ${\bf\Omega}_{\bf k}$~\cite{xiao2010berry,shen2017topological}.
To proceed, we consider a single-frequency driven electric field ${\bf E}(t) = E_{a}(t){\bf e}_{a} = {\rm Re}(\mathcal{E}_{a}e^{i\omega t}){\bf e}_{a}$ along the $a$ direction and write down the coefficients of non-equilibrium distribution to the first and second orders:
\begin{equation}
f_{1}=f_{1}^{(\omega)}e^{i\omega t}, \quad
f_{2}=f_{2}^{(0)}+f_{2}^{(2\omega)}e^{2 i\omega t}.\label{eq:f1f2}
\end{equation}
Since the second-order coefficient $f_2$ comprises both $\mathcal{E}_{a}^{2}$ and $\mathcal{E}_{a}^{2}e^{i2\omega t}$ terms, it contains both a constant and a frequency-doubled term. Specifically, by substituting Eq.~\eqref{eq:f1f2} into Eq.~\eqref{eq:JE}, one can have
\begin{equation}\label{eq:J}
J_{b}(\mathcal{E}_{a})={\rm Re}(J_{b}^{0}+J_{b}^{\omega}e^{i\omega t} + J_{b}^{2\omega}e^{i2\omega t}),
\end{equation} where
$J_{b}^{0}\!=\!\chi_{baa}^{(0)}\mathcal{E}_{a}^{2},~J_{b}^{\omega}\!=\!\chi_{ba}^{(1)}\mathcal{E}_{a},~J_{b}^{2\omega}\!=\!\chi_{baa}^{(2)}\mathcal{E}_{a}^{2}$, with coefficients given at the leading order by
\begin{eqnarray}
\chi_{ba}^{(1)}&\approx& -\varepsilon^{bac}\sigma_{c}^{}, \label{eq:chi_ba}\\
\chi_{baa}^{(0)}&\approx&\chi_{baa}^{(2)}\approx\frac{e^3\tau\varepsilon^{bac}D_{ac}}{2\hbar^2(1+i\omega\tau)} \label{eq:chi_baa}.
\end{eqnarray}
In the above, we have only kept terms to leading order in $\frac{e\tau}{\hbar}\mathcal{E}_{a}a$ (with the lattice constant $a$), which is very small in relevant experiments~\cite{ma2019observation}. For the subleading contributions, see equations~\eqref{eq:chi_0} to \eqref{eq:chi_2} in the Methods Section \ref{Methods_1}.
In the above, $\varepsilon^{bac}$ is the Levi-Civita anti-symmetric tensor; $a$, $b$, and $c$ index the spatial coordinates $x$, $y$, and $z$. The two relevant physical quantities are the linear Hall conductance $\sigma_c$ and the nonlinear Hall response BCD ($D_{ac}$). Specializing to the 2D $x-y$ plane such that the index $c=z$, we write the Hall conductance as~\cite{kubo1957statistical_1,thouless1982quantized,shen2017topological}
\begin{equation}
\sigma_{H}^{}\!=\!\frac{e^2}{h}\frac{1}{2\pi}\sum_{n}\int d^{2}{\bf k}~\Omega_{{\bf k},z}^{(n)}f_{0},
\label{eq:Hall_0}
\end{equation}
where $\Omega_{{\bf k},z}^{(n)}$ is the Berry curvature corresponding to the $n$-th eigenstate. However, the main quantity of focus in this work is the BCD:
\begin{equation}
D_{ac}=-\sum_{n}\int\frac{d^{2}{\bf k}}{(2\pi)^2}(\partial_{k_a}\epsilon_{\bf k}^{(n)})\Omega_{{\bf k},c}^{(n)}\frac{\partial f_{0}}{\partial\epsilon_{\bf k}^{(n)}}.\label{eq:BCD_0}
\end{equation}
Its detailed derivation can be found in Methods Section \ref{Methods_1}. Here, the derivative of the equilibrium distribution function is given by $\partial f_{0}/\partial\epsilon_{\bf k}^{(n)}=\frac{-e^{\left(\epsilon_{\bf k}^{(n)}\!-\!E_{F}\right)/(k_{B}T)}}{\left[1+e^{\left(\epsilon_{\bf k}^{(n)}\!-\!E_{F}\right)/(k_{B}T)}\right]^{2}k_{B}T}$, where $T$ is the temperature and $k_B$ is the Boltzmann constant. The above integral for the BCD indicates that the nonlinear Hall response is predominantly governed by the states near the Fermi energy $E_F$~\cite{du2018band,du2019disorder,du2021nonlinear,chen2024nonlinear}. We can interpret $D_{ac}$ as the momentum-space integral over the Berry curvature $\Omega_{{\bf k},c}^{(n)}$ weighted by the dispersion $\partial_{k_a}\epsilon_{\bf k}^{(n)}$ and occupation gradient $\partial f_{0}/\partial\epsilon_{\bf k}^{(n)}$, summed over all eigenstates $n$. Since inversion symmetry breaking is essential for the non-vanishing Berry curvature, only systems with broken inversion symmetry can exhibit non-zero BCD and thus nonlinear (transverse) response~\cite{du2018band,du2019disorder,du2021nonlinear,chen2024nonlinear}.

\subsubsection{Two-component models}

The simplest models with nontrivial linear Hall and nonlinear BCD responses require at least two bands for nontrivial Berry curvature. A generic 2-component model is expressed as the ansatz
\begin{equation}
{\cal H}({\bf k})=h_{0}\sigma_{0}+\sum_{i=x,y,z}h_{i}\sigma_{i},
\label{eq:H}
\end{equation}
where $\sigma_{x,y,z}$ are the Pauli matrices, and $\sigma_{0}$ is the $2\times2$ identity matrix. The eigenenergies for its upper ($+$) and lower ($-$) bands are
\begin{equation}
\epsilon_{\mathbf{k}}^{(\pm)}\!=\!h_{0} \!\pm\! \sqrt{h_{x}^{2}\!+\!h_{y}^{2}\!+\!h_{z}^{2}},\label{eq:energy}
\end{equation}
with corresponding eigenvectors $\left|\psi^{(\pm)}\right\rangle$ ($\bold k$ subscript index omitted for brevity). The $z$ component of Berry curvature, which contributes to the Hall response in two-dimensional systems, is given by~\cite{kubo1957statistical_1,kubo1957statistical_2,thouless1982quantized,shen2017topological,qin2023light,qin2024kinked}
\begin{equation}
\Omega^{(\pm)}_{\mathbf{k},z}\!=\!-2\varepsilon_{zxy}\!\frac{{\rm Im}\!\left\langle \!\psi^{(\pm)}\!\left|\!\frac{\partial{\cal H}^{(F)}}{\partial k_{x}}\right|\!\psi^{(\mp)}\!\right\rangle\!\!\left\langle\!\psi^{(\mp)}\!\left|\!\frac{\partial\!{\cal H}^{(F)}}{\partial k_{y}}\!\right|\!\psi^{(\pm)}\!\right\rangle}{\left[\epsilon_{\mathbf{k}}^{(\pm)}-\epsilon_{\mathbf{k}}^{(\mp)}\right]^{2}}.\label{eq:BC}
\end{equation}

Notice that, compared to more realistic lattice models under circularly polarized light, a simple linearized model (for example, the type-I Weyl semimetal) with additional quadratic $k^{2}$ terms can lead to different physics~\cite{bharti2023massless}. For a type-I Weyl semimetal, the population at one Weyl node induced by the left circularly polarized light is the same as that induced at the other Weyl node by the right circularly polarized light~\cite{bharti2023massless}. However, once the non-linearity (near the Weyl nodes) of the band structure is taken into account, the excitations induced near one Weyl node by the left circularly polarized light will no longer overlap perfectly with that induced near the other Weyl node by right circularly polarized light, even though the key Hall contributions remain similar.

\subsubsection{Effective Floquet Hamiltonian from optical driving}

The paradigmatic model for describing nonlinear Hall materials, for instance, monolayer WTe$_2$~\cite{kang2019nonlinear,ma2019observation,ye2023control,xiao2020berry}, is the  two-dimensional tilted massive Dirac model~\cite{sodemann2015quantum,du2018band,chen2024nonlinear,muechler2016topological}, which form the theoretical basis in nonlinear Hall effect experiments~\cite{du2018band,ma2019observation}:
\begin{equation}
{\cal H}({\bf k})\!=\!t_{0}k_x\sigma_{0}\!+\!vk_y\sigma_{x}\!+\!\eta vk_x\sigma_{y}\!+\!(m\!-\!\alpha k^{2})\sigma_{z},\label{eq:H}
\end{equation}
where $\eta$ is allowed to take values of $\pm 1$ in principle (in our numerics, we set $\eta=-1$), and $t_{0}$, $v$, $m$, and $\alpha$ are parameters that can be empirically fitted. The tilted term $t_{0}k_{x}\sigma_{0}$ breaks the inversion symmetry and is key to triggering the nonlinear Hall effect.

We next discuss how optical driving can modify the effective Hamiltonian and consequently significantly and precisely enhance the BCD. In a generic setting, the optical electric field propagating along the $z$ direction can be expressed as ${\bf E}(t) = \partial{\bf A}(t)/\partial t = E_{0}(\cos(\tilde{\omega} t),\cos(\tilde{\omega} t + \varphi))$, where $E_{0}$ is the amplitude of the electric field and $\tilde{\omega}$ is the angular frequency of the light. The phase $\varphi$ controls the polarization: $\varphi=0$ introduces linear polarization, while $\varphi=\mp\pi/2$ introduces left- or right-handed circular polarization. Integrating, we can have ${\bf A}(t)={\bf A}(t+T)=\tilde{\omega}^{-1}E_{0}(\sin(\tilde{\omega} t),\sin(\tilde{\omega} t + \varphi))$ that is of period $T=2\pi/\tilde{\omega}$. Notice that the light frequency $\tilde{\omega}$ is much higher than the ultralow frequency $\omega$ (17.77 Hz~\cite{ma2019observation}) of an longitudinal alternating current (a.c.) $I^{\omega}_{x}$, which is used to induce the transverse nonlinear Hall effect in experiments~\cite{ma2019observation,du2021nonlinear}.

Under optical driving, the motion of electrons is governed by minimal substitution of the lattice momentum with the electromagnetic gauge field ${\bf A}(t)$.
Hence, the photon-dressed effective Hamiltonian is given by
\begin{equation}\label{eq:Ht}
{\cal H}({\bf k},t)={\cal H}\left({\bf k} - \frac{e}{\hbar}{\bf A}(t)\right).
\end{equation}
In our Floquet band engineering proposal, we are interested in the off-resonant regime where the central Floquet band is far away from other replicas, such that the high-frequency expansion is applicable~\cite{magnus1954exponential,blanes2009magnus,lee2018floquet}. As such, we set the driving optical frequency to the representative value $\hbar\tilde{\omega}=1$ eV~\cite{sie2019time} ($\tilde{\omega}\sim1.519\times10^{15}$ Hz), which is much larger than the bandwidth~\cite{qin2023light,qin2022light,qin2022phase}. 
Under periodic driving through $\bold A(t)$, the effective Floquet Hamiltonian~\cite{oka2009photovoltaic,lee2018floquet,qin2023light,qin2022light,qin2022phase} ${\cal H}^{(F)}({\bf k})=\frac{i}{T}\ln\left[\mathcal{T}e^{-i\int^T_0 {\cal H}({\bf k},t)dt}\right]$ is the effective static Hamiltonian with the effects of the periodic driving ``averaged'' over one period. In the high-frequency regime, a closed-form solution exists via the Magnus expansion~\cite{magnus1954exponential,blanes2009magnus,lee2018floquet}
\begin{equation}\label{eq:HF0}
{\cal H}^{(F)}({\bf k}) = {\cal H}_{0} + \sum_{n=1}^{\infty}\frac{[{\cal H}_{-n}, {\cal H}_{n}]}{n\hbar\tilde{\omega}}+{\cal O}(\tilde{\omega}^{-2}),
\end{equation}
where ${\cal H}_{n} = \frac{1}{T} \int_{0}^{T}{\cal H}({\bf k},t) e^{in\tilde{\omega} t}dt$ is the $n$-th time Fourier component of $\mathcal{H}(\bold k,t)$. For our tilted Dirac model [Eq.~\eqref{eq:H}], all but the $n=1$ commutators vanish, as shown in Methods Section \ref{Methods_2}, and the Floquet Hamiltonian takes the form
\begin{equation}
{\cal H}^{(F)}({\bf k})=h_{0}\sigma_{0}+\sum_{i=x,y,z}h_{i}^{(F)}\sigma_{i},
\label{eq:H_F}
\end{equation}
where $h_{0}=t_{0}k_{x}$ as before,
\begin{eqnarray}
h_{x}^{(F)}&\!=&vk_{y}\left(1-\eta\frac{2\alpha A_{0}^{2}\sin\varphi}{\hbar\tilde{\omega}}\right), \label{eq:hx_F}\\
h_{y}^{(F)}&\!=&\eta vk_{x}\left(1-\eta\frac{2\alpha A_{0}^{2}\sin\varphi}{\hbar\tilde{\omega}}\right), \label{eq:hy_F}\\
h_{z}^{(F)}&\!=&m-A_{0}^{2}\left(\alpha + \eta\frac{v^{2}\sin\varphi}{\hbar\tilde{\omega}}\right)-\alpha k^2, \label{eq:hz_F}
\end{eqnarray} and $A_{0}=eE_{0}/(\hbar\tilde{\omega})$. Specifically, the intensity of the employed light is below the damage threshold. The maximum value of light amplitude used in our work is $A_{0}\sim1.49$ nm$^{-1}$, which corresponds to the amplitude of the electric field $E_{0}=\hbar\tilde{\omega}A_{0}/e\sim1.49\times10^{9}$ V/m. This maximum value of $E_{0}$ is smaller than $3.0\times10^{9}$ V/m in Refs.~\cite{wang2018light,bao2022light}, and it is also smaller than the maximum value $1.06\times10^{10}$  V/m in Ref.~\cite{sentef2015theory}. The detailed derivations for Eqs.~\eqref{eq:H_F} to \eqref{eq:hz_F} can be found in Methods Section \ref{Methods_2}.
Explicitly, we see that the optical driving has introduced new contributions proportional to $A_0^2$, which acts as a rescaling of the effective mass $m$ (up to an overall rescaling of the Hamiltonian) and hence ultimately its nonlinear response properties. To have non-vanishing BCD, note that $t_0\neq 0$ is still required for inversion symmetry breaking (Methods Section \ref{Methods_3}), even though time-reversal symmetry is already broken for any values of $t_0$ and $A_0$ (Methods Section \ref{Methods_4}).
Note that in the later numerical calculations, the Hamiltonian~\eqref{eq:H_F} would be regularized into its corresponding tight-binding lattice Hamiltonian.
The tight-binding lattice model for the Floquet Hamiltonian is shown in Supplementary Note 1.

%%%%%%%%%%%%%%%%%%%%%%%%%%%%%%%%%%%%%%%%%%%%%%%%%%
\begin{figure*}[htpb]
\centering
\includegraphics[width=.8\textwidth]{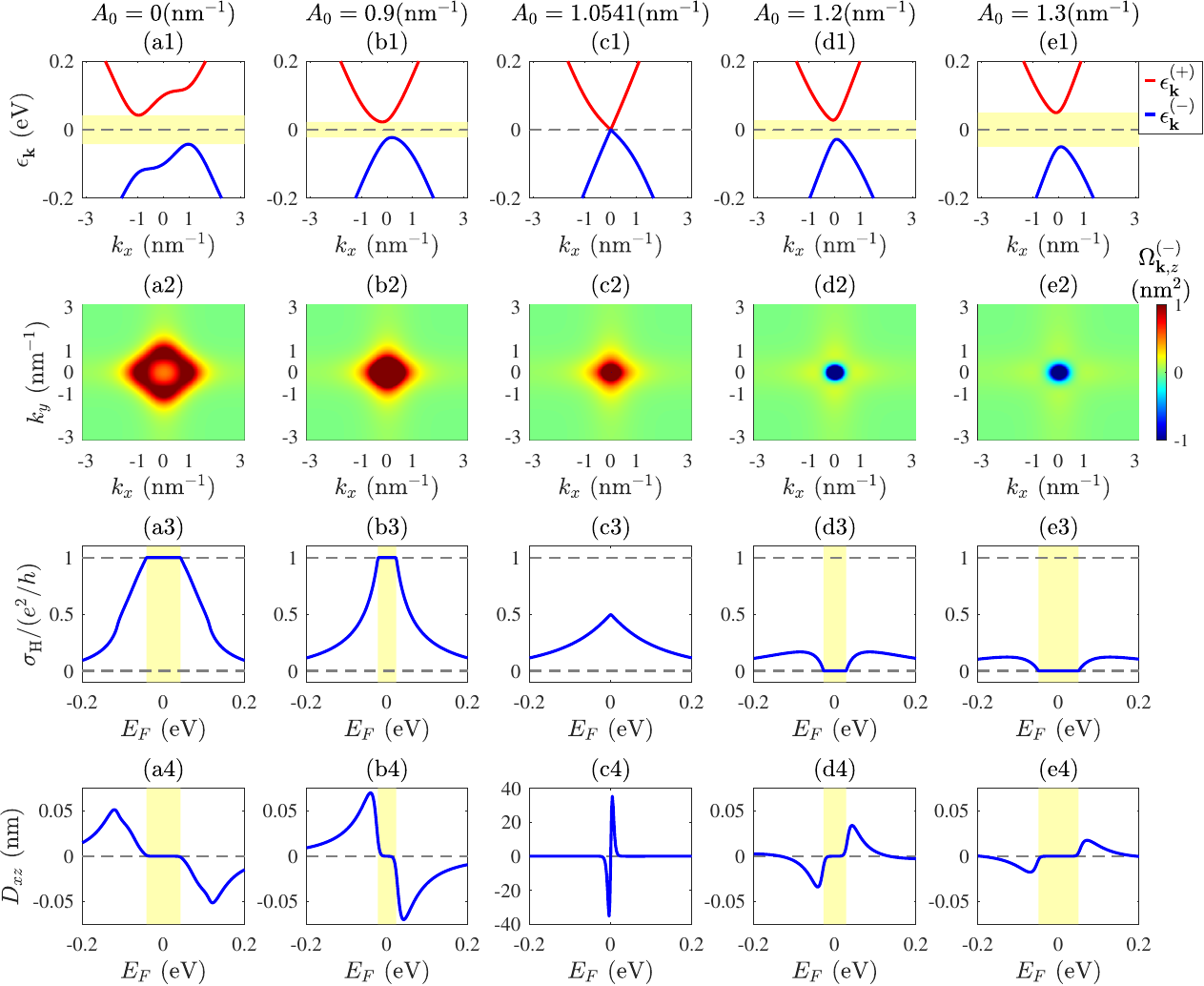}
\caption{\textbf{Light-induced topological band inversion and its large Berry curvature dipole (BCD) contribution.} Top row (a1-e1): The $k_y=0$ slice of the Floquet band structure [Eq.~\eqref{eq:energy}] for our nonlinear Hall medium [Eq.~\eqref{eq:H_F}], which exhibits a band inversion at laser amplitude $A_0\to A_{0c}\approx1.0541~\text{nm}^{-1}$. Second row (a2)-(e2): Berry curvature $\Omega^{(-)}_{\bold{k},z}$ [Eq.~\eqref{eq:BC}] of the lower band, which integrates to $+1$ for $A_0<A_{0c}$ and $0$ for $A_0>A_{0c}$. Third row (a3)-(e3): The resultant linear Hall conductance [Eq.~\eqref{eq:Hall_0}] as a function of the Fermi energy $E_F$, which is quantized at $+e^2/h$ in the gap (pale yellow) for $A_0<A_{0c}$ and $0$ for $A_0>A_{0c}$, corresponding to the Chern insulator (CI) and normal insulator (NI) phases, respectively. Fourth row (a4)-(e4): The nonlinear BCD $D_{xz}$ [Eq.~\eqref{eq:BCD_0}], which vanishes when $E_F$ is in the gap, but which peaks when $E_F$ is near the band edge. Near the transition at $A_0\approx A_{0c}$, the peak becomes drastically higher and narrower. The other parameters are $\varphi=\pi/2$ (right-handed circularly polarized light), $\hbar\tilde{\omega}=1$ eV, $t_{0}=0.05$ eV$\cdot$nm, $v=0.1$ eV$\cdot$nm, $\alpha=0.1$ eV$\cdot$nm$^2$, $m=0.1$ eV, $\eta=-1$, and $k_{B}T=0.003$ eV, i.e., $T\approx34.8136$ K, which are of the same order as those in Refs.~\cite{ma2019observation,muechler2016topological,liu2021short}.
}\label{fig:E_BC_C_BCD_phi05pi_together}
\end{figure*}
%%%%%%%%%%%%%%%%%%%%%%%%%%%%%%%%%%%%%%%%%%%%%%%%%%

To estimate the validity of the high-frequency expansion quantitatively, we evaluate the maximum instantaneous energy of the time-dependent Hamiltonian ${\cal H}\left({\bf k} - \frac{e}{\hbar}{\bf A}(t)\right)$ averaged over a Floquet period $\frac{1}{T}\int_{0}^{T}dt~\text{max}\left\{\big|\big|{\cal H}({\bf k},t)\big|\big|\right\}<\hbar\tilde{\omega}$ at the $\Gamma$ point ($k_x=k_y=0$), which gives the following constraint on the optical field parameters: ${\rm Max}\left(vA_{0},2\alpha A_{0}^{2}\right)<\hbar\tilde{\omega}$ i.e. $A_{0}<{\rm Min}\left(\frac{\hbar\tilde{\omega}}{v},\sqrt{\frac{\hbar\tilde{\omega}}{2\alpha}}\right)$.
In the high-frequency regime $\tilde{\omega} \sim 1.519\times10^{15}$ Hz ($\hbar\tilde{\omega}=1$ eV) with parameters set to $v=0.1$ eV$\cdot$nm and $\alpha=0.1$ eV$\cdot$nm$^2$, the same order as those in representative 2D Dirac materials such as WTe$_2$~\cite{kang2019nonlinear,ma2019observation,ye2023control,xu2018electrically,xiao2020berry}, MoTe$_2$~\cite{muechler2016topological}, and other WTe$_2$-type materials~\cite{liu2021short}, one can obtain $A_{0}=eE_{0}/(\hbar\tilde{\omega})<2.236$ nm$^{-1}$ ($E_{0}=A_{0}\hbar\tilde{\omega}/e<2.236\times10^{9}$ V/m), which corresponds to an incident light intensity~\cite{paschotta2016encyclopedia} of $I=\frac{1}{2}nc\varepsilon_{0}|E_{0}|^{2}<9.954\times10^{15}$ W/m$^2$, where $n$ is the refractive index, $c$ is the speed of light in vacuum, and
$\varepsilon_{0}$ is the vacuum permittivity. The refractive index $n\sim \mathcal{O}(1)$, with $n\approx1.5$~\cite{buchkov2021anisotropic} observed in monolayer WTe$_2$ in the deep-ultraviolet region.

\subsection{Divergent nonlinear Hall response near a topological transition}

\subsubsection{Light-induced topological transition}

Before showing how a large nonlinear Hall response can be achieved, we first elaborate on the topological phase transition induced by optical driving. Shown in Fig.~\ref{fig:E_BC_C_BCD_phi05pi_together} are the energy band structure, Berry curvature, Hall conductance, and BCD under right-handed circularly polarized light (i.e., $\varphi=\pi/2$) at different intensities $A_0$.

As can be theoretically predicted from Eqs.~\eqref{eq:hx_F} to \eqref{eq:hz_F}, the energy gap [pale yellow in Figs.~\ref{fig:E_BC_C_BCD_phi05pi_together}(a1)-\ref{fig:E_BC_C_BCD_phi05pi_together}(e1)] closes at the critical intensity $A_{0c}=\sqrt{m/\left(\alpha + \eta\frac{v^{2}\sin\varphi}{\hbar\tilde{\omega}}\right)}\approx1.0541$ nm$^{-1}$ [Fig.~\ref{fig:E_BC_C_BCD_phi05pi_together}(c1)]. Tuning the intensity across $A_{0c}$ not only closes and then opens the energy gap, but also gives rise to band inversion [Figs.~\ref{fig:E_BC_C_BCD_phi05pi_together}(a1)-\ref{fig:E_BC_C_BCD_phi05pi_together}(e1)], as evident from the change in sign of the Berry curvature of the lower band as shown in Figs.~\ref{fig:E_BC_C_BCD_phi05pi_together}(a2)-\ref{fig:E_BC_C_BCD_phi05pi_together}(e2), as analytically derived in Supplementary Note 2.

The fact that this band inversion corresponds to a topological transition can be seen in the change of the quantized value of the Hall conductance $\sigma_{H}^{}$ for $E_F$ in the gapped region (yellow) [Fig.~\ref{fig:E_BC_C_BCD_phi05pi_together}(a3)-\ref{fig:E_BC_C_BCD_phi05pi_together}(e3)]. When the light intensity $A_0$ is below the critical value $A_{0c}$, the energy spectrum is gapped [Figs.~\ref{fig:E_BC_C_BCD_phi05pi_together}(a1) and \ref{fig:E_BC_C_BCD_phi05pi_together}(b1)] and the corresponding Hall conductance {is quantized at $e^2/h$} within the gap as shown in Figs.~\ref{fig:E_BC_C_BCD_phi05pi_together}(a3) and \ref{fig:E_BC_C_BCD_phi05pi_together}(b3). This corresponds to the Chern insulator (CI) phase. When the light intensity $A_0>A_{0c}$, the spectrum is also gapped, as shown in Figs.~\ref{fig:E_BC_C_BCD_phi05pi_together}(d1) and \ref{fig:E_BC_C_BCD_phi05pi_together}(e1), but the corresponding Hall conductance equals zero as shown in Figs.~\ref{fig:E_BC_C_BCD_phi05pi_together}(d3) and \ref{fig:E_BC_C_BCD_phi05pi_together}(e3). This is the normal insulator (NI) phase.

Most saliently, at the topological phase transition $A_{0c}$, the BCD ($D_{xz}$) diverges for $E_F$ near the band edge, just outside of the pale yellow gap region. Very close to the transition, as shown in Fig.~\ref{fig:E_BC_C_BCD_phi05pi_together}(c4), it is orders of magnitude larger than that away from the transition, i.e., Figs.~\ref{fig:E_BC_C_BCD_phi05pi_together}(a4), \ref{fig:E_BC_C_BCD_phi05pi_together}(b4), \ref{fig:E_BC_C_BCD_phi05pi_together}(d4), and \ref{fig:E_BC_C_BCD_phi05pi_together}(e4). Since the BCD is proportional to the nonlinear Hall conductance, which can be directly measured, this divergence would have profound physical consequences, as explored in the following subsection. Analogous results under left-handed circularly polarized light ($\varphi=-\pi/2$) are qualitatively similar and can be found in Supplementary Note 3.

Notice that the circularly polarized light ($\varphi=\pm\pi/2$) can be used to break the time-reversal symmetry of the nonlinear Hall material (for example, WTe$_2$~\cite{du2018band,ma2019observation}) and realize a topological transition between normal and Chern insulator phases, but the linearly polarized light cannot break the time-reversal symmetry. Therefore, varying the amplitude of an incident circularly polarized laser drives a topological transition between normal and Chern insulator phases, and importantly allows the precise unlocking of nonlinear Hall currents comparable to or larger than the linear Hall contributions, notably without involving the tuning of any other physical parameter. Moreover, the high-harmonic generation is also a nonlinear process~\cite{bharti2022high}. As pointed out by Ref.~\cite{bharti2022high}, time-reversal symmetry-broken systems can lead to anomalous odd harmonics even based on a linearly polarized driving pulse. This is an alternate way to see the nonlinear Hall effect using a polarized driving light.

\subsubsection{Non-quantized current from nonlinear light-enhanced BCD}

%%%%%%%%%%%%%%%%%%%%%%%%%%%%%%%%%%%%%%%%%%%%%%%%%%
\begin{figure}[htpb]
\centering
\includegraphics[width=\columnwidth]{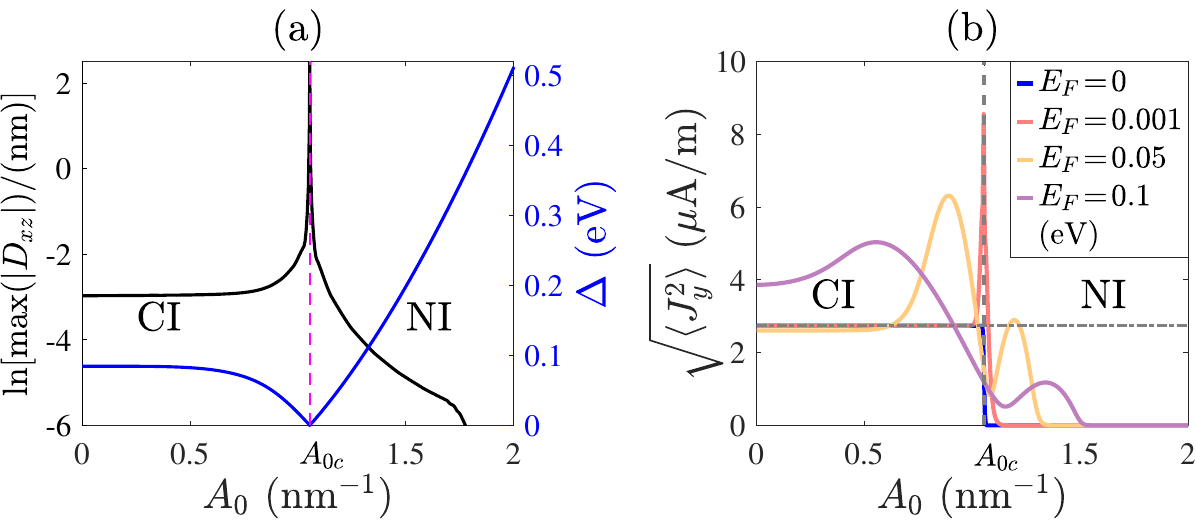}
\caption{
\textbf{Light enhanced Berry curvature dipole (BCD) and significant departure from linear Hall response.}
(a) BCD [Eq.~\eqref{eq:BCD_0}] peak max$(|D_{xz}|)$ (black) and band gap $\Delta={\rm min}[\epsilon_{\bf k}^{(+)}]-{\rm max}[\epsilon_{\bf k}^{(-)}]$ (blue) as a function of light amplitude $A_0$.  While the gap $\Delta$ vanishes as $A_0\to A_{0c}\approx1.0541$ nm$^{-1}$, leading to a transition between the Chern insulator (CI) and normal insulator (NI) phases, the BCD is dramatically enhanced. The BCD peak values correspond to the maximum of $D_{xz}$ over the broad window $E_F\in[-0.2,0.2]$ eV. (b) The root-mean-square Hall current density $\sqrt{\langle J_y^2\rangle}$ [Eq.~\eqref{eq:J_total}] as a function of $A_{0}$, from metallic ($E_F=0.1$ eV) to purely insulating ($E_F\to 0$) cases. While $\sqrt{\langle J_y^2\rangle}$ is understandably not quantized along the dashed line when the system is not insulating (purple, yellow), it is still not completely quantized even when $E_F$ is well within the gap (red). Instead, it exhibits a pronounced spike when $A_0$ is very close to $A_{0c}$ due to the large BCD contribution. Despite its narrowness, this spike is experimentally accessible due to the experimental ease of tuning $A_0$. Other than the electric field amplitude $\mathcal{E}_{x}$=0.1 V/m, $\tau\approx4.12434\times10^{-14}$ s~\cite{ma2019observation}, a.c. frequency $\omega=17.777$ Hz~\cite{ma2019observation}, and the other parameters used are identical to those in Fig.~\ref{fig:E_BC_C_BCD_phi05pi_together}.
} \label{fig:gap_BCD_phi05pi_together}
\end{figure}
%%%%%%%%%%%%%%%%%%%%%%%%%%%%%%%%%%%%%%%%%%%%%%%%%%

While it is commonly expected for the Hall current to be quantized according to the Chern number, in the presence of a large BCD, the Hall current should actually deviate considerably from its quantized value. In principle, this should be observable since a large BCD generally exists near a topological Chern transition with inversion symmetry breaking. However, in most realistic experimental settings, it is usually extremely difficult to tune the system to the BCD peak, which is extremely narrow, as plotted in black in Fig.~\ref{fig:gap_BCD_phi05pi_together}(a) for our system with $\varphi=\pi/2$ (right-handed circularly polarized light). A key advantage of our Floquet-induced BCD approach is that, by adjusting the laser intensity $A_0$, one is able to very precisely tune the system across the topological transition value $A_{0c}$, where the BCD (black) peaks and its corresponding bandgap $\Delta$ (blue) vanishes.

The tuning of the light amplitude to a precision of $A_{0}\sim0.1$ nm$^{-1}$ at least or better has been shown to be experimentally feasible. With fixed $\hbar\tilde{\omega}=1$ eV and the relation $A_{0}=eE_{0}/(\hbar\tilde{\omega})$, we can get the precision of light amplitude $A_{0}\sim0.025$ nm$^{-1}$ with $E_{0}=2.5\times10^{7}$ V/m~\cite{wang2013observation}, $A_{0}\sim0.116$ nm$^{-1}$ with $E_{0}=11.6\times10^{7}$ V/m~\cite{mahmood2016selective}, and $A_{0}\sim0.04$ nm$^{-1}$ with $E_{0}=4.0\times10^{7}$ V/m~\cite{mciver2020light}. Here $E_{0}$ is the amplitude of the electric field for the light. This precision in tuning $A_0$ allows one to observe the Hall current deviating greatly from its quantized value. Plotted in Fig.~\ref{fig:gap_BCD_phi05pi_together}(b) is the root-mean-square of the Hall current density $J_y$ [Eq.~\eqref{eq:J}] averaged over a period $\tilde{T}=2\pi/\omega$ as the driving light intensity $A_0$ is varied for an applied field $E_{x}^{\omega} = {\rm Re}(\mathcal{E}_{x}e^{i\omega t})$. Explicitly, from Eq.~\eqref{eq:J} to \eqref{eq:chi_baa},
\begin{eqnarray}
\langle J_{y}^{2}(\mathcal{E}_{x})\rangle&\!=\!&\int_{0}^{\tilde{T}}\!\frac{dt}{\tilde{T}}\!\left[J_{y}^{0} + J_{y}^{\omega}\cos(\omega t) + J_{y}^{2\omega}\cos(2\omega t) \right]^{2} \nonumber\\
&\!=\!&\left(J_{y}^{0}\right)^{2} + \frac{1}{2}\left(J_{y}^{\omega}\right)^{2} + \frac{1}{2}\left(J_{y}^{2\omega}\right)^{2} \nonumber\\
&\!=\!&\left[\chi_{yxx}^{(0)}\mathcal{E}_{x}^{2}\right]^{2} \!+\! \frac{1}{2}\left[\chi_{yx}^{(1)}\mathcal{E}_{x}\right]^{2} \!+\! \frac{1}{2}\left[\chi_{yxx}^{(2)}\mathcal{E}_{x}^{2}\right]^{2}\nonumber\\
&\approx &\,\frac{\mathcal{E}_x^2}{2}\left[\sigma_H^2 + \frac{3e^6\tau^2D_{xz}^2}{4\hbar^4}\mathcal{E}_x^2\right],
\label{eq:J_total}
\end{eqnarray}
where $\sigma_{H}^{}$ gives the linear Hall conductance contribution and the BCD ($D_{xz}$) provides the nonlinear contribution. When $E_F$ is outside of the bandgap (see Fig.~\ref{fig:E_BC_C_BCD_phi05pi_together}), $\sqrt{\langle J_{y}^{2}(\mathcal{E}_{x})\rangle}$ is not quantized as expected. However, for smaller Fermi energy, i.e., $E_F=0.001$ eV (red) where the system is clearly insulating, $\sqrt{\langle J_{y}^{2}(\mathcal{E}_{x})\rangle}$ does not remain quantized all the time; when $A_0$ is tuned very close to $A_{0c}$ where a Chern transition occurs, $\sqrt{\langle J_{y}^{2}(\mathcal{E}_{x})\rangle}$ exhibits a sharp spike due to the BCD ($D_{xz}$) peak. For a very narrow window of light intensity $A_0$, this nonlinear contribution can in fact be much larger than the usual linear contribution. Importantly, this window, albeit narrow, is readily experimentally accessible due to the ease of accurately tuning $A_0$~\cite{zhou2023pseudospin,kobayashi2023floquet}. Further discussion on the competition between the linear and nonlinear current contributions can be found in Methods Section \ref{Methods_5}.

\subsection{Enhanced BCD through Floquet quench}\label{6}

%%%%%%%%%%%%%%%%%%%%%%%%%%%%%%%%%%%%%%%%%%%%%%%%%%
\begin{figure}%[htpb]
\centering
\includegraphics[width=.5\textwidth]{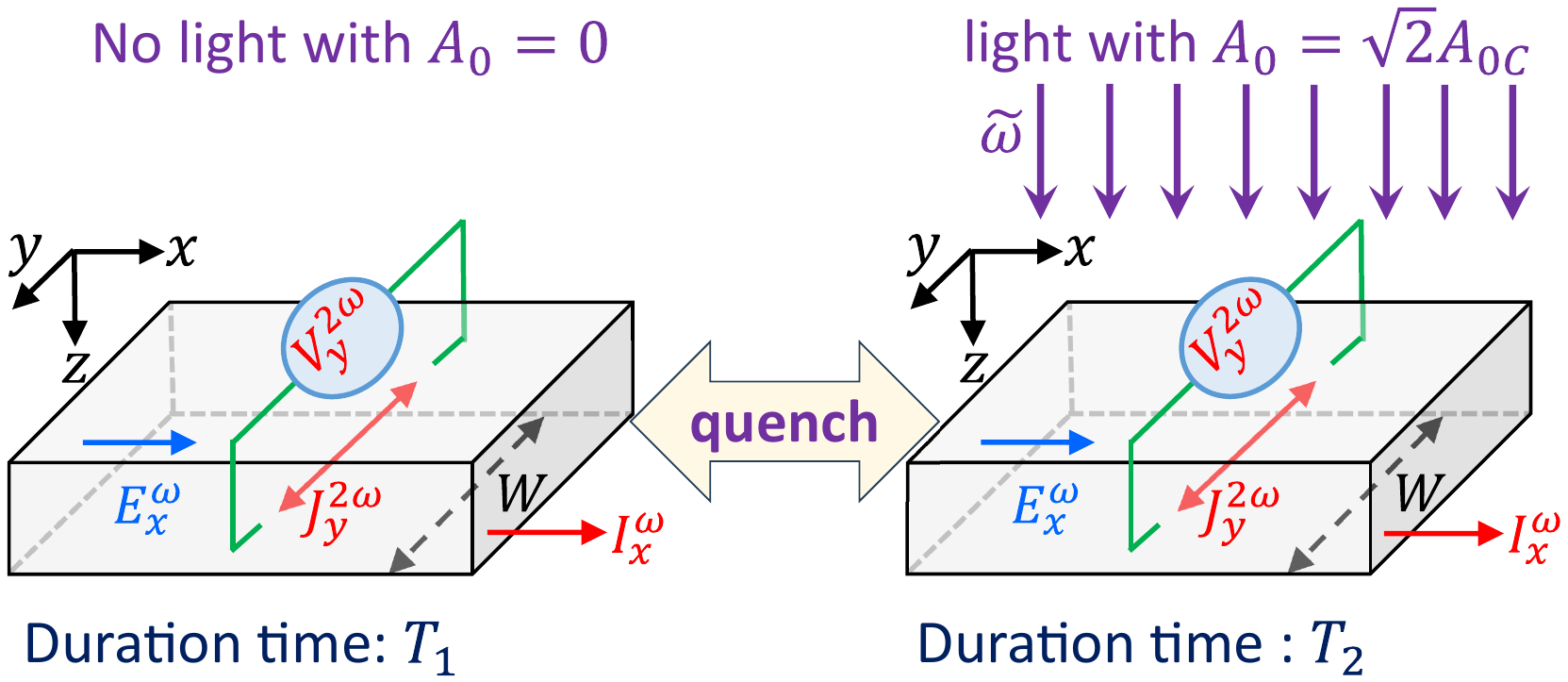}
\caption{\textbf{Proposed experimental setup for Floquet quench.}
In our nonlinear Hall material (taking the left figure for example), a lock-in amplifier is used to measure the nonlinear Hall voltage $V_{y}^{2\omega}$ resulting from a longitudinal electrical field $E_{x}^{\omega}={\rm Re}(\mathcal{E}_{x}e^{i\omega t})$ induced by an a.c. $I^\omega_x$ with ultralow frequency $\omega$, which is much lower than the optical driving frequency $\tilde{\omega}$. Left figure: In the duration time $T_{1}$, there is no illumination on the nonlinear Hall material. Right figure: In the duration time $T_{2}$, there is a high-frequency irradiated light (purple) with amplitude $A_{0}=\sqrt{2}A_{0c}$ and high frequency $\tilde{\omega}$ illuminating the nonlinear Hall material. The total periodic time is $T=T_{1}+T_{2}$.
}\label{fig:Schematic_nonlinear_quench}
\end{figure}
%%%%%%%%%%%%%%%%%%%%%%%%%%%%%%%%%%%%%%%%%%%%%%%%%%

An interesting extension of our above-mentioned approach involves performing a Floquet quench on the normal and Chern insulator phases. Since from Fig.~\ref{fig:E_BC_C_BCD_phi05pi_together}, the Chern insulator phase exists without optical driving and the normal insulator phase is generated by a strong polarized light, we shall periodically quench the polarized light to alternate rapidly between light-off and light-on (shown in Fig.~\ref{fig:Schematic_nonlinear_quench}), as can be achieved in experiments on attosecond pulses of light~\cite{castelvecchi2023physicists,paul2001observation,hentschel2001attosecond,mrudul2021high,kruchinin2018colloquium,ghimire2019high,nourbakhsh2021high}. Without loss of generality, we employ right-handed circularly polarized light with $\varphi=\pi/2$ in the following discussions.

%%%%%%%%%%%%%%%%%%%%%%%%%%%%%%%%%%%%%%%%%%%%%%%%%%
\begin{figure}%[htpb]
\centering
\includegraphics[width=.45\textwidth]{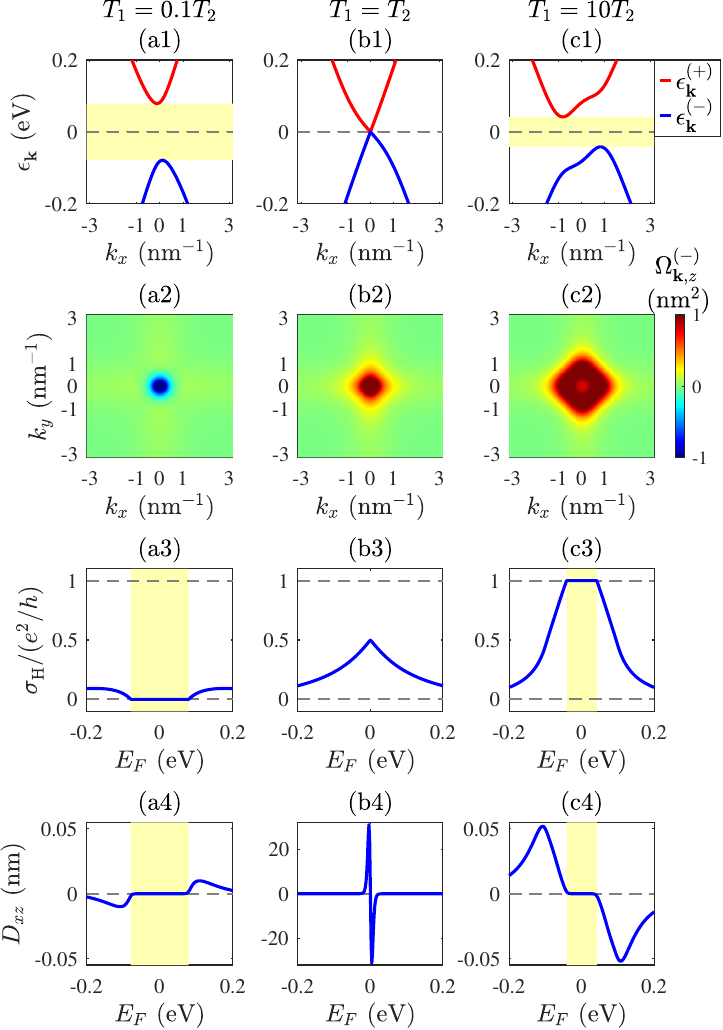}
\caption{\textbf{Quench-induced topological band inversion and Berry curvature dipole (BCD) peaks.}
(a1-c1) Floquet energy bands [Eq.~\eqref{eq:Heff}] for the $k_{y}=0$ slice. In (a1) where $T_{1}=0.1T_{2}$ and (c1) where $T_2=0.1T_1$, the ${\cal H}_{2}({\bf k})$ with light amplitude $A_0=\sqrt{2}A_{0c}$ and ${\cal H}_{1}({\bf k})$ with $A_0=0$ are respectively dominant. They respectively correspond to the normal insulator (NI) and Chern insulator (CI) phases and are both gapped (pale yellow). But in (b1) where $T_{1}=T_{2}$, nontrivial contributions from ${\cal H}_{1}({\bf k})$ and ${\cal H}_{2}({\bf k})$ cancel each other out, leaving a gapless band structure. (a2-c2) Berry curvature [Eq.~\eqref{eq:BC}] of the lower band, whose integral over the $k_x$-$k_y$ plane interpolates between the quantized values of $0$ and $+1$ as $T_1/T_2$ increases. (a3-c3) The corresponding Hall conductance [Eq.~\eqref{eq:Hall_0}], which exhibits these quantized values when $E_{F}$ is within the band gap (pale yellow). (a4-c4) The BCD ($D_{xz}$) [Eq.~\eqref{eq:BCD_0}], which vanishes for $E_F$ within the band gap but which peaks at the band edges, particularly when $T_1=T_2$. The other parameters are identical to those in Fig.~\ref{fig:E_BC_C_BCD_phi05pi_together}. Here, $T_{2}$ is fixed at $0.1\hbar/$eV$\approx6.58212\times10^{-17}$ s.} \label{fig:quench_E_BC_C_BCD_phi05pi_together}
\end{figure}
%%%%%%%%%%%%%%%%%%%%%%%%%%%%%%%%%%%%%%%%%%%%%%%%%%

As shown in Fig.~\ref{fig:Schematic_nonlinear_quench}, we consider a periodic two-step quench with a total period of $T_{1}+T_{2}$, such that each period consists of an odd (even) step governed by the Hamiltonian under light-off (light-on) polarized light described by ${\cal H}_{1}({\bf k})$ [${\cal H}_{2}({\bf k})$], for a duration of $T_{1}$ [$T_{2}$].  ${\cal H}_{1}({\bf k})$ denotes the Hamiltonian without light, i.e., $A_{0}=0$, and ${\cal H}_{2}({\bf k})$ denotes the Hamiltonian under right-handed circularly polarized light with $\varphi=\pi/2$ and $A_{0}=\sqrt{2}A_{0c}=\sqrt{2m/\left(\alpha + \eta\frac{v^{2}\sin\varphi}{\hbar\tilde{\omega}}\right)}\approx1.49071$ nm$^{-1}$, which from Fig.~\ref{fig:E_BC_C_BCD_phi05pi_together} is well within the normal insulating phase. The effective Floquet Hamiltonian is given by~\cite{qin2023light}
\begin{equation}
{\cal H}_\text{eff}({\bf k})\equiv\frac{i}{T_{1}+T_{2}}\ln[e^{-i{\cal H}_{2}({\bf k})T_{2}}e^{-i{\cal H}_{1}({\bf k})T_{1}}],\label{eq:Heff}
\end{equation}
whose gap (pale yellow in Fig.~\ref{fig:quench_E_BC_C_BCD_phi05pi_together}) depends intimately on the values of both $T_1$ and $T_2$. For our choice of light amplitudes $A_0=0$ for ${\cal H}_{1}({\bf k})$ and $A_0=\sqrt{2}A_{0c}$ for ${\cal H}_{2}({\bf k})$, the gap closes at $T_1=T_2$ [Fig.~\ref{fig:quench_E_BC_C_BCD_phi05pi_together}(b1)], which also induces a band inversion that separates the CI phase (with $T_1>T_2$) from the NI phase (with $T_1<T_2$). This is shown through the Berry curvature of the lower band $\Omega_{{\bf k},z}^{(-)}$ in Fig.~\ref{fig:quench_E_BC_C_BCD_phi05pi_together}(a2-c2) and its corresponding linear Hall conductance in Fig.~\ref{fig:quench_E_BC_C_BCD_phi05pi_together}(a3-c3). Similar to that in Fig.~\ref{fig:E_BC_C_BCD_phi05pi_together}, the BCD ($D_{xz}$) also exhibits a sharp peak near the band edge when the band gap vanishes, although this behavior is now precisely tunable through the ratio $T_1/T_2$ (with $T_2$ fixed at $0.1\hbar/eV\approx 6.58212\times 10^{-17}$s), rather than the laser amplitude $A_0$. Analogous results for left-handed circularly polarized optical driving ($\varphi=-\pi/2$) can be found in the Supplementary Note 3.

%%%%%%%%%%%%%%%%%%%%%%%%%%%%%%%%%%%%%%%%%%%%%%%%%%
\begin{figure}%[htpb]
\centering
\includegraphics[width=\columnwidth]{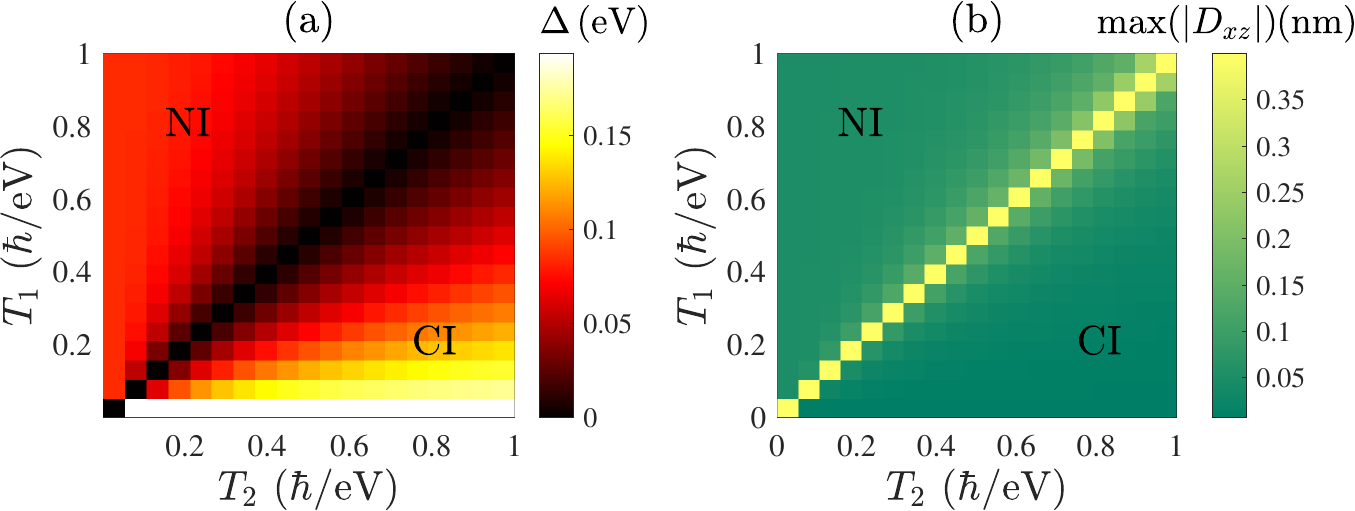}
\caption{\textbf{Tunable band gap and Berry curvature dipole (BCD) peaks from Floquet quench.}
Our quenching protocol offers versatility in the tuning of the nonlinear Hall response through the two independent quench durations $T_1$ and $T_2$. (a) The band gap $\Delta$ [Eq.~\eqref{eq:Heff}] in the $(T_2,T_1)$ parameter space. (b) The quench-enhanced BCD peak [Eq.~\eqref{eq:BCD_0}], which saliently peaks around the $T_{1}=T_{2}$ line where the gap closes. Here NI denotes the normal insulator phase and CI denotes the Chern insulator phase. The other parameters are identical to those in Fig.~\ref{fig:E_BC_C_BCD_phi05pi_together}.
}\label{fig:quench_gap_BCD_max_3D_together}
\end{figure}
%%%%%%%%%%%%%%%%%%%%%%%%%%%%%%%%%%%%%%%%%%%%%%%%%%

With independent control over both $T_1$ and $T_2$ step durations, this Floquet quench approach offers even greater versatility in tuning the topological phase transition and approaching the BCD peak. Shown in Fig.~\ref{fig:quench_gap_BCD_max_3D_together} are the band gap $\Delta$ and BCD peak $\text{max}|D_{xz}|$ in the parameter spaces of $T_1$ and $T_2$. Evidently, the phase boundary occurs along the $T_1=T_2$ line, which also corresponds to the peaks in the BCD. By simultaneously adjusting $T_1$ and $T_2$, maximal nonlinear Hall response from the BCD term can be obtained.

\subsection{Measurement of the nonlinear BCD response}

\begin{figure}%[htpb]
\centering
\includegraphics[width=\columnwidth]{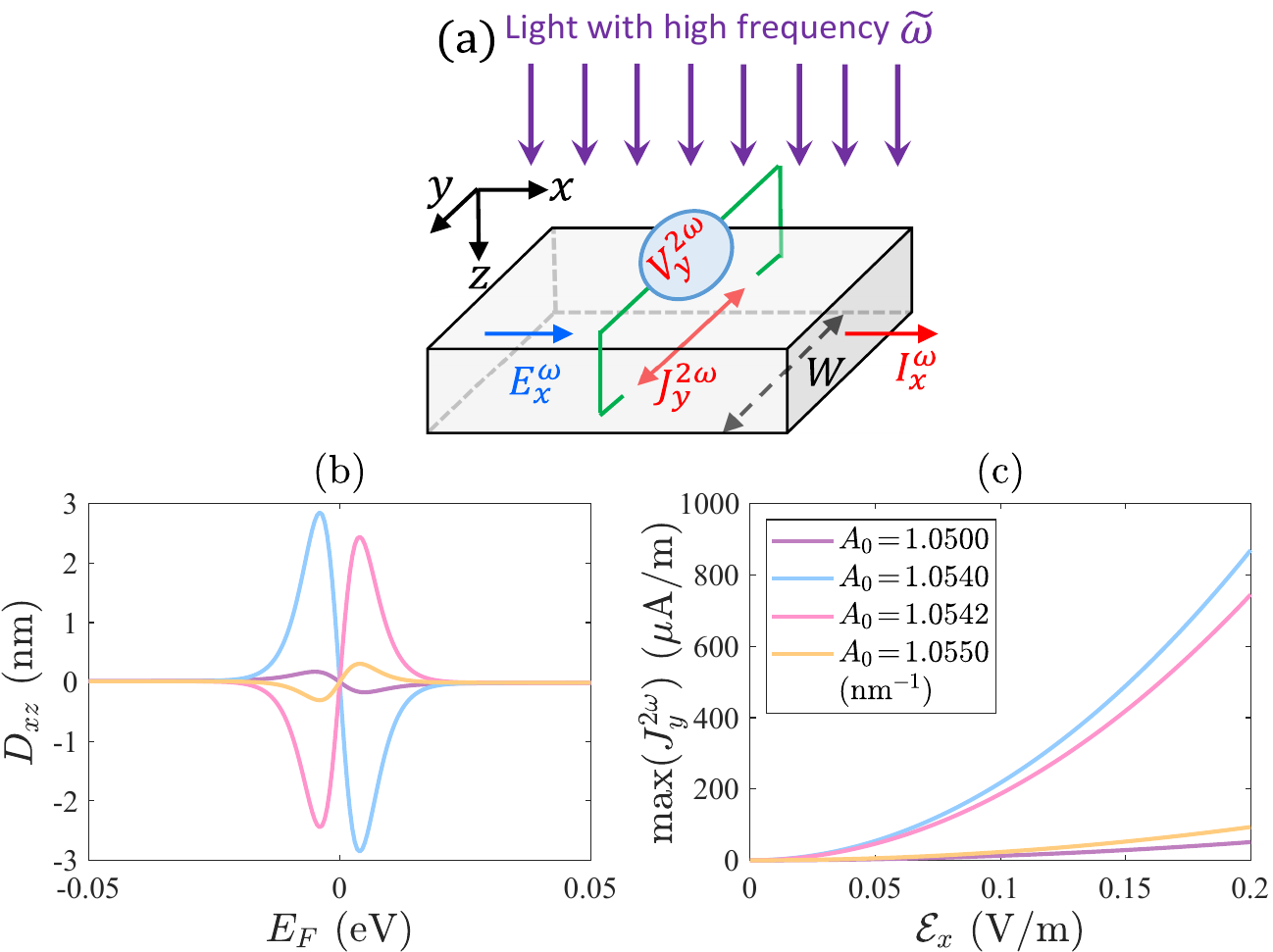}
\caption{\textbf{Proposed experimental setup and signatures of nonlinear Hall response.}
(a) In our nonlinear Hall insulator, a lock-in amplifier can be used to measure the nonlinear Hall voltage $V_y^{2\omega}$ resulting from a longitudinal electrical field $E_{x}^{\omega}={\rm Re}(\mathcal{E}_{x}e^{i\omega t})$ induced by an a.c. $I^\omega_x$ of ultralow frequency $\omega$, which is much lower than the optical driving frequency $\tilde{\omega}$. High-frequency ($\tilde{\omega}$) irradiated light (purple) of amplitude $A_0\approx A_{0c}$ produces the requisite Floquet-enhanced Berry curvature dipole (BCD). (b) The BCD ($D_{xz}$) [Eq.~\eqref{eq:BCD_0}] as a function of the Fermi energy $E_F$ for various light amplitudes $A_0$ close to the critical value $A_{0c}\approx1.0541$ nm$^{-1}$ where a topological transition occurs, as labeled in (c). The BCD is dramatically enhanced nearest to $A_{0c}$, peaking at $\sim3$ nm that is comparable to experimentally measured values in~\cite{ma2019observation}. (c) The peak of the second-harmonic Hall current density $J_{y}^{2\omega}=\chi_{yxx}^{(2)}\mathcal{E}_{x}^{2}$ [Eq.~\eqref{eq:J}], which varies nonlinearly with the amplitude of the applied perpendicular electric field $\mathcal{E}_{x}$, and is greatly enhanced for $A_0$ nearest to $A_{0c}$. The other parameters are $\varphi=\pi/2$ (right-handed circularly polarized light), $\hbar\tilde{\omega}=1$ eV, $t_{0}=0.05$ eV$\cdot$nm, $v=0.1$ eV$\cdot$nm, $\alpha=0.1$ eV$\cdot$nm$^2$, $m=0.1$ eV [Eq.~\eqref{eq:H}], $k_{B}T=0.003$ eV, i.e., $T\approx34.8$ K, $\tau\approx4.124\times10^{-14}$ s~\cite{ma2019observation}, and $\omega=17.77$ Hz~\cite{ma2019observation}.
}\label{fig:BCD_Jy2_max_phi05pi_together}
\end{figure}

The nonlinear response due to light-enhanced BCD can be experimentally measured in the schematic setup shown in Fig.~\ref{fig:BCD_Jy2_max_phi05pi_together}(a). Floquet driving is achieved through high-frequency laser pumping (purple), whose amplitude $A_0$ can be tuned to sensitively adjust the BCD strength. The nonlinear response current density $J_{y}^{2\omega}$, which is perpendicular to the longitudinal electric field $E_{x}^{\omega}={\rm Re}(\mathcal{E}_{x}e^{i\omega t})$ induced by an alternating current (a.c.) $I_{x}^{\omega}$ at ultralow frequency $\omega$, can be measured with a lock-in amplifier on its corresponding potential difference $V_{y}^{2\omega}$.

%%%%%%%%%%%%%%%%%%%%%%%%%%%%%%%%%%%%%%%%%%%%%%%%%%
%%%%%%%%%%%%%%%%%%%%%%%%%%%%%%%%%%%%%%%%%%%%%%%%%%

To obtain the BCD ($D_{xz}$) explicitly, we invoke the nonlinear Hall response relation $J_{y}^{2\omega}\!=\!\chi_{yxx}^{(2)}\mathcal{E}_{x}^2$ in Eq.~\eqref{eq:chi_baa}:
\begin{equation}
D_{xz}=\frac{2\hbar^{2}(1+i\omega\tau)J_{y}^{2\omega}}{e^3\tau\varepsilon^{yxx}\mathcal{E}_{x}^2},
\label{eq:D_bd_1}
\end{equation} in which $\mathcal{E}_{x}$ and $J_{y}^{2\omega}$ can be obtained from the excitation current $I_x^\omega$ and measured perpendicular $V_y^{2\omega}$ via Ohm\rq{}s law as $\mathcal{E}_{x}=\frac{I_{x}^{\omega}}{\sigma W}$ and $J_{y}^{2\omega}=\frac{\sigma V_{y}^{2\omega}}{W}$. As shown in Fig.~\ref{fig:BCD_Jy2_max_phi05pi_together}(a), $W$ is the effective sample width, and the longitudinal conductance is of the order of $\sigma\sim 10^{-5}$ S~\cite{ma2019observation}. From the Drude formula $\sigma=\frac{n_{0}e^{2}\tau}{\tilde{m}}$, one can also estimate the relaxation time $\tau$ to be $\sim 10^{-14}$ s with the effective mass being $\tilde{m}=0.3m_{e}$, $m_e$ the electron mass~\cite{ma2019observation}, and $n_{0}$ the charge density.
Tuning this charge density $n_0$ also modifies the Fermi energy $E_F$, as demonstrated by first-principle calculations in a related system~\cite{ma2019observation}.

Experimentally, it suffices to use an ultralow a.c. frequency $\omega$ of the order of $10-10^2$ Hz~\cite{ma2019observation}, such that $\omega\tau\to 0$ and we obtain
\begin{equation}
D_{xz}=\frac{2\hbar^{2}\sigma^{3}V_{y}^{2\omega}W}{e^3\tau(I_{x}^{\omega})^{2}}.\label{eq:D_bd_2}
\end{equation}
Based on these physical parameters, the BCD ($D_{xz}$) is plotted in Fig.~\ref{fig:BCD_Jy2_max_phi05pi_together}(b) for various optical driving amplitudes $A_0$ near the critical value $A_{0c}\approx1.0541~\text{nm}^{-1}$. It exhibits peaks of up to $3$ nm for $A_0=1.0540~\text{nm}^{-1}$ [blue curve in Fig.~\ref{fig:BCD_Jy2_max_phi05pi_together}(b)], the same order as in the experimentally measured BCD in bilayer WTe$_2$~\cite{ma2019observation}, even though our system is monolayer. The peak width increases with temperature and is about $E_F\sim 10^{-2}$ eV when calculated with the temperature of $T\approx 34.8$ K taken from the experiment in~\cite{ma2019observation}.

As evident, $D_{xz}$ exhibits high sensitivity to the value of $A_0$ near the critical value $A_{0c}$. This also translates to the high sensitivity of the peak of the nonlinear Hall current density $J_{y}^{2\omega}$, as plotted in Fig.~\ref{fig:BCD_Jy2_max_phi05pi_together}(c). The nonlinearity of the current response is evident from the curvature of max($J_{y}^{2\omega}$) plotted against the perpendicular electric field amplitude $\mathcal{E}_x$, particularly for values of $A_0$ closest to $A_{0c}$ (blue, pink).

Notice that there are two frequency scales that differ by several orders of magnitude in this work: the very slow driving frequency of the longitudinal electrical field ($\omega=17.77$ Hz~\cite{ma2019observation}), and the rapid driving from the irradiation ($\hbar\tilde{\omega}=1$ eV). As shown in Fig.~\ref{fig:BCD_Jy2_max_phi05pi_together}(a), the driving frequency of the longitudinal electrical field (blue) ${\bf E}$ (i.e., $E^{\omega}_{x}$) in Eq.~\eqref{eq:JE} is $\omega=17.77$ Hz~\cite{ma2019observation}, which is very low. Therefore, the semiclassical current [Eq.~\eqref{eq:JE}], which involves the Berry curvature, is valid. On the other hand, the high frequency $\hbar\tilde{\omega}=1$ eV is for our irradiated light (purple) as shown in Fig.~\ref{fig:BCD_Jy2_max_phi05pi_together}(a). Here, the rapid driving modifies the band structure and does not participate in the semiclassical theory, which is based on the effective Floquet model after ``integrating out\rq\rq{} the rapid driving. Furthermore, our irradiated light (purple) is along $z$ direction as shown in Fig.~\ref{fig:BCD_Jy2_max_phi05pi_together}(a) and Eq.~\eqref{eq:BC} only provides the $z$ component of the Berry curvature. Therefore, the irradiated light cannot contribute to the semiclassical current [Eq.~\eqref{eq:JE}] with an additional term $\frac{e}{\hbar}{\bf E}_{\rm light}\times{\bf\Omega}_{\bf k}$, where ${\bf E}_{\rm light}$ is the electrical field of our irradiated light and ${\bf E}_{\rm light}\times{\bf\Omega}_{\bf k}=E_{\rm light}\Omega_{{\bf k},z}({\bf e}_{z}\times{\bf e}_{z})=0$. But the irradiated light can modify the Berry curvature through the electromagnetic gauge field [Eq.~\eqref{eq:Ht}] within the Floquet theory as shown in Eq.~\eqref{eq:BC}.

\subsection{Conclusion}

We find that nonlinear Hall materials can exhibit a strong light-enhanced Berry curvature dipole (BCD) and hence a nonlinear Hall response when excited by circularly polarized lasers. This was established using the two-dimensional tilted massive Dirac Hamiltonian that accurately models known nonlinear Hall materials such as WTe$_2$~\cite{kang2019nonlinear,ma2019observation,ye2023control,xu2018electrically,xiao2020berry}, MoTe$_2$~\cite{lai2021third,muechler2016topological}, and other WTe$_2$-type materials~\cite{liu2021short}. More generally, however, the dramatic light-induced BCD enhancement is expected to occur in all media with simultaneously broken time-reversal and inversion symmetries, albeit only in close proximity to a topological transition.

The generically very narrow window of large BCD places significant challenges on its experimental observation. This challenge is crucially addressed through our Floquet approach, which enables convenient, precise access to the topological transition by varying the laser amplitude or quench duration. We find a significant light-enhanced peak for the BCD at a specific critical intensity of incident light, at which there is a light-induced topological band inversion: when the light intensity is subcritical, the system is a Chern insulator, whereas exceeding the critical intensity results in a transition to a normal insulator phase. Conversely, by measuring the BCD peak as a function of light intensity, the location of the topological phase transition point can also be accurately determined. In all, our approach not only precisely assesses a topological transition and its accompanying BCD peak, but also enhances the sensitivity and reproducibility of nonlinear Hall measurements in the larger realm of nonlinear physics~\cite{he2019nonlinear,tuloup2020nonlinearity}.

Particularly, we find that our work and Ref.~\cite{aversa1995nonlinear} focus on different physics that is distinguished by the magnitude of the frequency of the pumping light. The previous work~\cite{aversa1995nonlinear} focuses on the nonlinear optical susceptibilities of semiconductors. Notice that the optical susceptibilities require that the frequency of the driving light be in the same order of magnitudes as the bandwidth. It is called a resonant light when its frequency is about the same as the bandwidth. Therefore, the electrons can transition between the conduction band and the valence band, driven by the resonant light. However, in our work, we focus on the off-resonant light, whose frequency is much larger than the bandwidth. In this case, the electrons cannot transition between the conduction band and the valence band, driven by the off-resonant light. Therefore, an effective Floquet Hamiltonian based on a high-frequency expansion method provided in our work is applicable for this off-resonant light situation.

\begin{acknowledgements}
We acknowledge helpful discussions with Hao-Jie Lin and Xiao-Bin Qiang.
C.H.L. and F.Q. acknowledge support from the QEP2.0 Grant from the Singapore National Research Foundation (Grant No. NRF2021-QEP2-02-P09) and the Singapore Ministry of Education Academic Research Fund  Tier-II Grant (Award No. MOE-T2EP50222-0003). R.C. acknowledges the support from the National Natural Science Foundation of China (Grant No. 12304195) and the Chutian Scholars Program in Hubei Province.
\end{acknowledgements}

\section{Data availability}
No datasets were generated or analyzed during the current study.

\section{Author contributions}
F.Q. performed theoretical calculations and wrote parts of the manuscript. R.C. assisted in improving the theoretical analysis and numerical calculations. C.H.L. supervised the project, wrote the manuscript and provided technical guidance. All authors discussed the analytical and numerical results and contributed to all aspects of the manuscript.

\section{Competing interests}
The authors declare no competing interests.

%\bibliography{references_nonlinear_TB}
%%%%%%%%%%%%%%%%%%%%%%  References %%%%%%%%%%%%%%%%%%%%%%
%\bibliographystyle{apsrev4-1}
%\bibliographystyle{plain}
%\bibliography{references}
%\setcounter{equation}{0}
%\setcounter{figure}{0}
%\setcounter{table}{0}
%\setcounter{page}{1}
%\setcounter{section}{0}
%\renewcommand{\theequation}{S\arabic{equation}}
%\renewcommand{\thefigure}{S\arabic{figure}}
%\renewcommand{\thesection}{S\arabic{section}}
%\renewcommand{\thepage}{S\arabic{page}}
%
%\onecolumngrid
%\flushbottom
%\newpage
%\appendix
%\setcounter{equation}{0}
%\setcounter{figure}{0}
%\setcounter{table}{0}
%\setcounter{page}{1}
%\setcounter{section}{0}
%\setcounter{secnumdepth}{3}
%\renewcommand{\theequation}{S\arabic{equation}}
%\renewcommand{\thefigure}{S\arabic{figure}}
%\renewcommand{\thesection}{S\Roman{section}}
%\renewcommand{\thepage}{S\arabic{page}}
%\renewcommand*{\citenumfont}[1]{S#1}
%\renewcommand*{\bibnumfmt}[1]{[S#1]}
\onecolumngrid
\flushbottom
%\newpage
\begin{sloppypar}

\section{METHODS}

\subsection{Nonlinear Hall effect}\label{Methods_1}

In this subsection, we will introduce the nonlinear Hall effect, where the transverse electric current contains both linear and quadratic (or even higher) contributions from the external electric field due to higher-order Berry curvature corrections~\cite{gao2014field,sodemann2015quantum,du2018band,du2019disorder,du2021nonlinear,chen2024nonlinear,ma2019observation,lai2021third}.

In the quantum Hall (or quantum anomalous Hall) effect, under broken time-reversal symmetry, longitudinal current does not flow due to the band gap. Instead, due to topological in-gap states, a transverse linear Hall current density $J_{y}=\sigma_{xy}E_{x}$ exists at the same a.c. frequency. It is well known that at zero temperature, a perturbative expansion in linear response theory gives the Hall conductance $\sigma_{xy}=-\frac{e^{2}2\pi}{h}\int\frac{d^2{\bf k}}{(2\pi)^2}\varepsilon^{xyz}\Omega_{{\bf k},z}$, where $\varepsilon^{xyz}$ is the Levi-Civita anti-symmetric tensor. But the nonlinear Hall response tells of another story: Instead of the usual Ohm's law, we have a quadratic $I$-$V$ relation, i.e., the transverse current density is proportional to the second power of the longitudinal electric field, due to high-order Berry curvature-like contributions. Generically, we can write the response to the electric current density as
\begin{equation}\label{eq:Appendix_J}
J_{i}=\sigma_{ij}E_{j}+\chi_{ijk}E_{j}E_{k}+\xi_{ijkl}E_{j}E_{k}E_{l}+\cdots,
\end{equation} where $\{i,j,k,l\}\in\{x,y,z\}$, $\mathbf{E}=(E_{x},E_{y},E_{z})$ is the external electric field; the first term is the linear term with frequency as the external electric field; and the second term is the leading-order nonlinear transverse response with a doubled a.c. frequency that can be measured with a lock-in amplifier.
In the following subsubsections, we derive and study the coefficient $\chi_{ijk}$ for the nonlinear Hall effect.

\subsubsection{Equations of motion}

Under an electric field ${\bf E}$, the equations of motion of semi-classical electronic wavepackets are given by~\cite{chang1995berry,sundaram1999wave,xiao2010berry,shen2017topological}
\begin{eqnarray}
\dot{\bf r}&=&\frac{1}{\hbar}\nabla_{\bf k}\epsilon_{\bf k}^{} - \dot{\bf k}\times{\bf\Omega}_{\bf k},\label{eq:r}\\
\dot{\bf k}&=&-\frac{e}{\hbar}{\bf E},\label{eq:kt}
\end{eqnarray} where both the position ${\bf r}$ and wave vector ${\bf k}$ simultaneously characterize the center-of-mass and momentum of a wavepacket, $\dot{\bf r}$ and $\dot{\bf k}$ are their time derivatives, ${\bf E}$ is the external longitudinal electric field, $\epsilon_{\bf k}^{}$ is the energy band dispersion, and ${\bf\Omega}_{\bf k}$ is the Berry curvature defined in ${\bf k}$ space~\cite{xiao2010berry,shen2017topological}.

Substituting Eq.~(\ref{eq:kt}) into Eq.~(\ref{eq:r}), the formula of $\dot{\bf r}$ becomes
\begin{equation}
\tilde{\bf v}_{\bf k}^{}=\dot{\bf r}={\bf v}_{\bf k} + \frac{e}{\hbar}{\bf E}\times{\bf\Omega}_{\bf k},\label{eq:rt_B0}
\end{equation} where we have defined the wavepacket velocity as~\cite{xiao2010berry}
\begin{equation}\label{eq:vk}
{\bf v}_{\bf k}^{}=\frac{1}{\hbar}\nabla_{\bf k}\epsilon_{\bf k}^{}.
\end{equation}

\subsubsection{Boltzmann equation within the relaxation time approximation}

In the case of thermal equilibrium without an external field, the distribution function is the Fermi-Dirac distribution function
\begin{equation}
f_{0}(\epsilon_{\bf k})=\frac{1}{e^{\beta(\epsilon_{\bf k}^{}-\mu)} + 1},
\end{equation} where $\beta=1/(k_{B}T)$ with Boltzmann constant $k_B$ and temperature $T$, and $\mu$ is the chemical potential.

When collisions exist, we consider the non-equilibrium distribution function $f({\bf r},{\bf k},t)$
\begin{equation}
\frac{df({\bf r},{\bf k},t)}{dt}=\left(\frac{\partial f}{\partial t}\right)_{\rm coll}.
\end{equation}
Further, we expand the non-equilibrium distribution as
\begin{equation}
f({\bf r},{\bf k},t)=f({\bf r}-d{\bf r},{\bf k}-d{\bf k},t-dt)=f({\bf r}-\dot{\bf r}dt,{\bf k}-\dot{\bf k}dt,t-dt) + \left(\frac{\partial f}{\partial t}\right)_{\rm coll}dt,
\end{equation}
and obtain
\begin{equation}
f({\bf r}-\dot{\bf r}dt,{\bf k}-\dot{\bf k}dt,t-dt) = f({\bf r},{\bf k},t) - \dot{\bf r}\cdot\frac{\partial f}{\partial{\bf r}}dt - \dot{\bf k}\cdot\frac{\partial f}{\partial{\bf k}}dt - \frac{\partial f}{\partial t}dt + {\cal O}(dt^2).
\end{equation}
Thus, the Boltzmann equation acts as
\begin{equation}\label{eq:Boltzmann}
\dot{\bf r}\cdot\frac{\partial f}{\partial{\bf r}} + \dot{\bf k}\cdot\frac{\partial f}{\partial{\bf k}} + \frac{\partial f}{\partial t} = \left(\frac{\partial f}{\partial t}\right)_{\rm coll},
\end{equation}
where the left side is a drift term and the right side is a collision term.

Under an external electric field ${\bf E}$, the drift term of the Boltzmann equation [Eq.~(\ref{eq:Boltzmann})] becomes
\begin{equation}\label{eq:drift}
\frac{\partial f}{\partial t} + \dot{\bf r}\cdot\frac{\partial f}{\partial{\bf r}} + \dot{\bf k}\cdot\frac{\partial f}{\partial{\bf k}}
=\frac{\partial f}{\partial t} - \frac{e}{\hbar}{\bf E}\cdot\frac{\partial f}{\partial{\bf k}},
\end{equation} where we have used $\frac{\partial f}{\partial{\bf r}}=0$ since the field is spatially homogeneous on the length scale of a wavepacket.
We employ a simple relaxation time approximation
\begin{equation}\label{eq:coll}
\left(\frac{\partial f}{\partial t}\right)_{\rm coll}\approx
-\frac{f-f_0}{\tau}.
\end{equation}
Then, the Boltzmann equation~(\ref{eq:Boltzmann}) becomes
\begin{equation}\label{eq:Boltzmann_f1}
f-f_0=-\tau\frac{\partial f}{\partial t} + \frac{e\tau}{\hbar}{\bf E}\cdot\frac{\partial f}{\partial{\bf k}} 
=-\tau\frac{\partial f}{\partial t} + e\tau{\bf E} \cdot\frac{1}{\hbar}\nabla_{\bf k}\epsilon_{\bf k}^{}\frac{\partial f}{\partial\epsilon_{\bf k}^{}} =
-\tau\frac{\partial f}{\partial t} + e\tau{\bf E}\cdot{\bf v}_{\bf k}^{}\frac{\partial f}{\partial\epsilon_{\bf k}^{}},
\end{equation} where ${\bf v}_{\bf k}^{}=\frac{1}{\hbar}\nabla_{\bf k}\epsilon_{\bf k}^{}$.

Furthermore, the Boltzmann equation~(\ref{eq:Boltzmann_f1}) becomes
\begin{equation}\label{eq:Boltzmann_f1_B0}
f-f_0=-\tau\frac{\partial f}{\partial t} + e\tau{\bf v}_{\bf k}\frac{\partial f}{\partial\epsilon_{\bf k}^{}}\cdot{\bf E}
=-\tau\frac{\partial f}{\partial t} + \frac{e\tau}{\hbar}{\bf E}\cdot\frac{\partial f}{\partial{\bf k}}
=-\tau\frac{\partial f}{\partial t} + \frac{e\tau}{\hbar}E_{a}\frac{\partial f}{\partial k_a}.
\end{equation}
Importantly, this allows for a controlled approximation of the distribution function as
\begin{equation}
f=\frac{f_0}{1+\tau\partial_{t}-\frac{e\tau}{\hbar}E_{a}\partial_{k_a}}
\approx\sum_{n=0}^{\infty}\left(-\tau\partial_{t}+\frac{e\tau}{\hbar}E_{a}\partial_{k_a} \right)^{n}f_{0}
=f_{0}+f_{1}+f_{2}+\cdots\approx f_{0}+f_{1}+f_{2},
\end{equation} where $f_n$ refers to the contribution to $f$ that is of the $n$-order in the field $\bold E$ (and also of $e\tau/\hbar$). Here, we noted that $\partial_{t}f_0=0$.

We specialize in an oscillating (i.e., a.c.) electric field, ${\bf E}(t)=E_{a}^{\omega}(t){\bf e}_{a}=\frac{1}{2}{\rm Re}(\mathcal{E}_{a}e^{i\omega t}+\mathcal{E}^{*}_{a}e^{-i\omega t}){\bf e}_{a}={\rm Re}(\mathcal{E}_{a}e^{i\omega t}){\bf e}_{a}$, where the driving field oscillates harmonically in time, but it is uniform in space. With the amplitude vector $\mathcal{E}_a$ and frequency $\omega$, we have the linear contribution $f_1$~\cite{du2019disorder}
\begin{eqnarray}
f_{1}&=&\sum_{n=0}^{\infty}(-\tau\partial_{t})^{n}\frac{e\tau}{\hbar}{\rm Re}(\mathcal{E}_{a} e^{i\omega t})\partial_{k_a}f_{0}
=\frac{e\tau}{\hbar}\sum_{n=0}^{\infty}(-\tau\partial_{t})^{n}{\rm Re}(\mathcal{E}_{a} e^{i\omega t})\partial_{k_a}f_{0} \nonumber\\
&=&\frac{e\tau}{\hbar}\sum_{n=0}^{\infty}{\rm Re}[\mathcal{E}_{a}e^{i\omega t}(-i\omega\tau)^{n} ]\partial_{k_a}f_{0}
=\frac{e\tau}{\hbar}{\rm Re}\left(\frac{\mathcal{E}_{a}e^{i\omega t}}{1+i\omega\tau} \right)\partial_{k_a}f_{0},
\end{eqnarray} where we noted that
\begin{equation}
\sum_{n=1}^{\infty}(-\tau\partial_{t})^{n-1}\frac{e\tau}{\hbar}E_{a}^{\omega}\partial_{k_a}=\frac{e\tau}{\hbar}\sum_{n'=0}^{\infty}(-\tau\partial_{t})^{n'}E_{a}^{\omega}\partial_{k_a}\propto E_{a}^{\omega}
\end{equation} is still linear in the electric field $E_{a}^{\omega}$. Furthermore, the leading-order nonlinear response contribution is given by
\begin{eqnarray}
f_{2}&=&\sum_{n=0}^{\infty}(-\tau\partial_{t})^{n}\frac{e\tau}{2\hbar}{\rm Re}(\mathcal{E}_{a}e^{i\omega t}+\mathcal{E}^{*}_{a}e^{-i\omega t})\partial_{k_a}\left[\sum_{m=0}^{\infty}(-\tau\partial_{t})^{m}\frac{e\tau}{2\hbar}{\rm Re}(\mathcal{E}_{a}e^{i\omega t}+\mathcal{E}^{*}_{a}e^{-i\omega t})\partial_{k_a}\right]f_{0} \nonumber\\
&=&\frac{e^{2}\tau^{2}}{4\hbar^2}\sum_{n=0}^{\infty}(-\tau\partial_{t})^{n}\left\{{\rm Re}(\mathcal{E}_{a}e^{i\omega t}+\mathcal{E}^{*}_{a}e^{-i\omega t})\partial_{k_a}\left[{\rm Re}\left(\frac{\mathcal{E}_{a}e^{i\omega t}}{1+i\omega\tau} + \frac{\mathcal{E}^{*}_{a}e^{-i\omega t}}{1-i\omega\tau} \right)\partial_{k_a}\right]f_{0}\right\} \nonumber\\
&=&\frac{e^{2}\tau^{2}}{4\hbar^2}\sum_{n=0}^{\infty}(-\tau\partial_{t})^{n}{\rm Re}\left[\left(\frac{\mathcal{E}_{a}\mathcal{E}_{a}e^{i2\omega t}}{1+i\omega\tau} + \frac{\mathcal{E}_{a}^{*}\mathcal{E}_{a}^{*}e^{-i2\omega t}}{1-i\omega\tau} \right) + \left(\frac{\mathcal{E}_{a}\mathcal{E}_{a}^{*}}{1-i\omega\tau} + \frac{\mathcal{E}_{a}^{*}\mathcal{E}_{a}}{1+i\omega\tau} \right)\right]\partial_{k_a}\partial_{k_a}f_{0} \nonumber\\
&=&\frac{e^{2}\tau^{2}}{4\hbar^2}\sum_{n=0}^{\infty}{\rm Re}\left[\left(\frac{\mathcal{E}_{a}\mathcal{E}_{a}e^{i2\omega t}(-i2\omega\tau)^{n}}{1+i\omega\tau} + \frac{\mathcal{E}_{a}^{*}\mathcal{E}_{a}^{*}e^{-i2\omega t}(i2\omega\tau)^{n}}{1-i\omega\tau} \right) + \left(\frac{\mathcal{E}_{a}\mathcal{E}_{a}^{*}}{1-i\omega\tau} + \frac{\mathcal{E}_{a}^{*}\mathcal{E}_{a}}{1+i\omega\tau} \right)\right]\partial_{k_a}\partial_{k_a}f_{0} \nonumber\\
&=&\frac{e^{2}\tau^{2}}{4\hbar^2}{\rm Re}\left[\left(\frac{\mathcal{E}_{a}\mathcal{E}_{a}e^{i2\omega t}}{(1+i\omega\tau)(1+i2\omega\tau)} + \frac{\mathcal{E}_{a}^{*}\mathcal{E}_{a}^{*}e^{-i2\omega t}}{(1-i\omega\tau)(1-i2\omega\tau)} \right) + \left(\frac{\mathcal{E}_{a}^{*}\mathcal{E}_{a}}{1+i\omega\tau} + \frac{\mathcal{E}_{a}\mathcal{E}_{a}^{*}}{1-i\omega\tau} \right)\right]\partial_{k_a}\partial_{k_a}f_{0} \nonumber\\
&=&\frac{e^{2}\tau^2}{2\hbar^2}{\rm Re}\left[\frac{\mathcal{E}_{a}\mathcal{E}_{a}e^{i2\omega t}}{(1+i\omega\tau)(1+i2\omega\tau)}  + \frac{\mathcal{E}_{a}^{*}\mathcal{E}_{a}}{1+i\omega\tau} \right]\partial_{k_a}\partial_{k_a}f_{0},
\end{eqnarray} where $\tau$ is taken to be spatially constant, i.e., $\partial_{k_a}\tau=0$. Note that $f_2$ contains two distinct types of nonlinear contributions, one oscillating at the doubled frequency $2\omega$ and the other static.

\subsubsection{Electric current density}

Explicitly, from the definition of the non-equilibrium distribution function, we have the electric current density
\begin{equation}\label{eq:J_E}
{\bf J}({\bf E})=\frac{-e}{V}\sum_{\bf k}\dot{\bf r}f_{\bf k}\,\approx\,\frac{-e}{V}\sum_{\bf k}\dot{\bf r}(f_{0}+f_{1}+f_{2}),
\end{equation} where $V=L^{D}$ is the volume in $D$ dimensions. As $\bold k$ lives in the lattice Brillouin zone, we should replace the sum over ${\bf k}$ by the integral
$\sum_{\bf k}\rightarrow V\int\frac{d^{D}{\bf k}}{(2\pi)^D}$. Substituting Eqs.~(\ref{eq:rt_B0}) and (\ref{eq:Boltzmann_f1_B0}) into (\ref{eq:J_E}), we obtain, to the leading nonlinear order,
\begin{equation}\label{eq:J_E_B0}
{\bf J}({\bf E})\approx\frac{-e}{V}\sum_{\bf k}\dot{\bf r}(f_{0}+f_{1}+f_{2})
=\frac{-e}{V}\sum_{\bf k}\left({\bf v}_{\bf k} + \frac{e}{\hbar}{\bf E}\times{\bf\Omega}_{\bf k}\right)(f_{0}+f_{1}+f_{2}).
\end{equation}
Further, we have the $b$-component in the electric current density
\begin{eqnarray}\label{eq:Ja_E_B0}
J_{b}
&\approx&\frac{-e}{V}\sum_{\bf k}\left(v_{k_b} + \frac{e}{\hbar}\varepsilon^{bac}E_{a}^{\omega}\Omega_{{\bf k},c}\right)(f_{0}+f_{1}+f_{2}) \nonumber\\
&=&\frac{-e}{V}\sum_{\bf k}\left[v_{k_b} + \frac{e}{2\hbar}\varepsilon^{bac}{\rm Re}(\mathcal{E}_{a}e^{i\omega t}+\mathcal{E}_{a}^{*}e^{-i\omega t})\Omega_{{\bf k},c}\right] \nonumber\\
&&\times\left\{f_{0} + \frac{e\tau}{2\hbar}{\rm Re}\left(\frac{\mathcal{E}_{a}e^{i\omega t}}{1+i\omega\tau} + \frac{\mathcal{E}_{a}^{*}e^{-i\omega t}}{1-i\omega\tau} \right)\partial_{k_a}f_{0} + \frac{e^2\tau^2}{4\hbar^2}{\rm Re}\left(\frac{\mathcal{E}_{a}^{*}\mathcal{E}_{a}}{1+i\omega\tau} + \frac{\mathcal{E}_{a}\mathcal{E}_{a}^{*}}{1-i\omega\tau} \right)\partial_{k_a}\partial_{k_a}f_{0} \right. \nonumber\\
&&\left.+ \frac{e^2\tau^2}{4\hbar^2}{\rm Re}\left[\frac{\mathcal{E}_{a}\mathcal{E}_{a}e^{i2\omega t}}{(1+i\omega\tau)(1+i2\omega\tau)} + \frac{\mathcal{E}_{a}^{*}\mathcal{E}_{a}^{*}e^{-i2\omega t}}{(1-i\omega\tau)(1-i2\omega\tau)} \right]\partial_{k_a}\partial_{k_a}f_{0} \right\} \nonumber\\
&=&\frac{-e}{V}\sum_{\bf k}\left[v_{k_b} + \frac{e}{\hbar}\varepsilon^{bac}{\rm Re}(\mathcal{E}_{a}e^{i\omega t})\Omega_{{\bf k},c}\right] \nonumber\\
&&\times\left\{f_{0} + \frac{e\tau}{\hbar}{\rm Re}\left(\frac{\mathcal{E}_{a}e^{i\omega t}}{1+i\omega\tau} \right)\partial_{k_a}f_{0} + \frac{e^2\tau^2}{2\hbar^2}{\rm Re}\left(\frac{\mathcal{E}_{a}^{*}\mathcal{E}_{a}}{1+i\omega\tau} \right)\partial_{k_a}\partial_{k_a}f_{0} + \frac{e^2\tau^2}{2\hbar^2}{\rm Re}\left[\frac{\mathcal{E}_{a}\mathcal{E}_{a}e^{i2\omega t}}{(1+i\omega\tau)(1+i2\omega\tau)} \right]\partial_{k_a}\partial_{k_a}f_{0} \right\} \nonumber\\
&\equiv&{\rm Re}(J_{b}^{0} + J_{b}^{\omega}e^{i\omega t} + J_{b}^{2\omega}e^{i2\omega t}),
\end{eqnarray} where $\varepsilon^{abc}$ is the Levi-Civita anti-symmetric tensor; $a$, $b$, and $c$ stand for the spatial coordinates $x$, $y$, and $z$,
\begin{eqnarray}
{\rm Re}(J_{b}^{0})&=&\frac{e^2\tau}{4\hbar^2}\frac{(-e)}{V}\sum_{\bf k}\left[\varepsilon^{bac}{\rm Re}\left(\frac{\mathcal{E}_{a}\mathcal{E}_{a}^{*}}{1+i\omega\tau} + \frac{\mathcal{E}_{a}^{*}\mathcal{E}_{a}}{1-i\omega\tau} \right)\Omega_{{\bf k},c}\partial_{k_a} + \tau v_{k_b}{\rm Re}\left(\frac{\mathcal{E}_{a}\mathcal{E}_{a}^{*}}{1+i\omega\tau} + \frac{\mathcal{E}_{a}^{*}\mathcal{E}_{a}}{1-i\omega\tau} \right)\partial_{k_a}\partial_{k_a} \right]f_{0} \nonumber\\
&=&\frac{e^2\tau}{4\hbar^2}\frac{(-e)}{V}\sum_{\bf k}\left[\varepsilon^{bac}{\rm Re}\left(\frac{\mathcal{E}_{a}\mathcal{E}_{a}^{*}}{1+i\omega\tau} + \frac{\mathcal{E}_{a}^{*}\mathcal{E}_{a}}{1-i\omega\tau} \right)\Omega_{{\bf k},c}\partial_{k_a} + \tau v_{k_b}{\rm Re}\left(\frac{\mathcal{E}_{a}\mathcal{E}_{a}^{*}}{1+i\omega\tau} + \frac{\mathcal{E}_{a}^{*}\mathcal{E}_{a}}{1-i\omega\tau} \right)\partial_{k_a}\partial_{k_a} \right]f_{0} \nonumber\\
&=&\frac{e^2\tau}{4\hbar^2}\frac{(-e)}{V}\sum_{\bf k}{\rm Re}\left(\frac{\mathcal{E}_{a}\mathcal{E}_{a}^{*}}{1+i\omega\tau} + \frac{\mathcal{E}_{a}^{*}\mathcal{E}_{a}}{1-i\omega\tau} \right)\left(\varepsilon^{bac}\Omega_{{\bf k},c}\partial_{k_a} + \tau v_{k_b}\partial_{k_a}\partial_{k_a} \right)f_{0} \nonumber\\
&=&\frac{e^2\tau}{2\hbar^2}\frac{(-e)}{V}\sum_{\bf k}{\rm Re}\left(\frac{\mathcal{E}_{a}\mathcal{E}_{a}^{*}}{1+i\omega\tau}\right)\left(\varepsilon^{bac}\Omega_{{\bf k},c}\partial_{k_a} + \tau v_{k_b}\partial_{k_a}\partial_{k_a} \right)f_{0},\\
{\rm Re}(J_{b}^{\omega})
&=&\frac{e}{\hbar}\frac{(-e)}{V}\sum_{\bf k}\left[\varepsilon^{bac}\mathcal{E}_{a}\Omega_{{\bf k},c} + {\rm Re}\left(\frac{\tau v_{k_b}\mathcal{E}_{a}}{1+i\omega\tau}\right)\partial_{k_a} \right]f_{0} \nonumber\\
&=&\frac{e\mathcal{E}_{a}}{\hbar}\frac{(-e)}{V}\sum_{\bf k}\left[\varepsilon^{bac}\Omega_{{\bf k},c} + {\rm Re}\left(\frac{\tau v_{k_b}}{1+i\omega\tau}\right)\partial_{k_a}\right]f_{0}, \\
{\rm Re}(J_{b}^{2\omega})
&=&\frac{e^2\tau}{2\hbar^2}\frac{(-e)}{V}\sum_{\bf k}\left\{{\rm Re}\left(\frac{\varepsilon^{bac}\mathcal{E}_{a}\mathcal{E}_{a}}{1+i\omega\tau}\right)\Omega_{{\bf k},c}\partial_{k_a} + {\rm Re}\left[\frac{\tau v_{k_b}\mathcal{E}_{a}\mathcal{E}_{a}}{(1+i\omega\tau)(1+i2\omega\tau)}\right]\partial_{k_a}\partial_{k_a}\right\}f_{0} \nonumber\\
&=&{\rm Re}\left[\frac{e^2\tau}{2\hbar^2(1+i\omega\tau)}\right]\frac{(-e)}{V}\sum_{\bf k}\left[{\rm Re}(\varepsilon^{bac}\mathcal{E}_{a}\mathcal{E}_{a})\Omega_{{\bf k},c}\partial_{k_a} + {\rm Re}\left(\frac{\tau v_{k_b}\mathcal{E}_{a}\mathcal{E}_{a}}{1+i2\omega\tau}\right)\partial_{k_a}\partial_{k_a}\right]f_{0} \nonumber\\
&=&{\rm Re}\left[\frac{e^2\tau\mathcal{E}_{a}\mathcal{E}_{a}}{2\hbar^2(1+i\omega\tau)}\frac{(-e)}{V}\sum_{\bf k}\left(\varepsilon^{bac}\Omega_{{\bf k},c}\partial_{k_a} + \frac{\tau v_{k_b}}{1+i2\omega\tau}\partial_{k_a}\partial_{k_a}\right)f_{0}\right].
\end{eqnarray}

By defining
\begin{equation}
J_{b}^{0}=\chi_{baa}^{(0)}\mathcal{E}_{a}\mathcal{E}_{a}^{*},~~J_{b}^{\omega}=\chi_{ba}^{(1)}\mathcal{E}_{a},~~J_{b}^{2\omega}=\chi_{baa}^{(2)}\mathcal{E}_{a}\mathcal{E}_{a},
\end{equation} where we identify the linear Hall coefficient $\chi_{ba}^{(1)}$ and nonlinear Hall coefficients $\chi_{ba}^{(0)},\chi_{ba}^{(2)}$ as
\begin{eqnarray}
\chi_{baa}^{(0)}&=&\frac{e^3\tau}{2\hbar^2(1+i\omega\tau)}\left(\varepsilon^{bac}D_{ac} + \tau U_{baa} \right)\approx \frac{e^3\tau\varepsilon^{bac}D_{ac}}{2\hbar^2(1+i\omega\tau)},\label{eq:chi_0}\\
\chi_{ba}^{(1)}&=&-\frac{e^2}{\hbar}\left[\varepsilon^{bac}A_{c} + \frac{\tau}{(1+i\omega\tau)}C_{ba} \right]\approx -\frac{e^2}{\hbar}\varepsilon^{bac}A_{c} = -\varepsilon^{bac}\sigma_{c}^{},\label{eq:chi_1}\\
\chi_{baa}^{(2)}&=&\frac{e^3\tau}{2\hbar^2(1+i\omega\tau)}\left[\varepsilon^{bac}D_{ac} + \frac{\tau}{(1+i2\omega\tau)}U_{baa} \right]\approx \frac{e^3\tau\varepsilon^{bac}D_{ac}}{2\hbar^2(1+i\omega\tau)},\label{eq:chi_2}
\end{eqnarray} where ${\bf v}_{\bf k}^{}=\frac{1}{\hbar}\nabla_{\bf k}\epsilon_{\bf k}^{}$, $v_{a}=\frac{1}{\hbar}\partial_{k_a}\epsilon_{\bf k}^{}$, and
\begin{eqnarray}
A_{c}^{}&\!=\!&\sum_{n}\int\frac{d^{D}{\bf k}}{(2\pi)^D}\Omega_{{\bf k},c}^{(n)}f_{0},\\
\sigma_{c}^{}&\!=\!&\frac{e^{2}2\pi}{h}A_{c}^{}\!=\!\frac{e^{2}2\pi}{h}\sum_{n}\int\frac{d^{D}{\bf k}}{(2\pi)^D}\Omega_{{\bf k},c}^{(n)}f_{0},\\
C_{ba}&\!=\!&\frac{2\pi}{h}\sum_{n}\int\frac{d^{D}{\bf k}}{(2\pi)^D}(\partial_{k_b}\epsilon_{\bf k}^{(n)})(\partial_{k_a} f_{0}),\\
D_{ac}&\!=\!&-\sum_{n}\int\frac{d^{D}{\bf k}}{(2\pi)^D}(\partial_{k_a}\epsilon_{\bf k}^{(n)})\Omega_{{\bf k},c}^{(n)}\frac{\partial f_{0}}{\partial\epsilon_{\bf k}^{(n)}},\\
U_{baa}&\!=\!&-\frac{2\pi}{h}\sum_{n}\int\frac{d^{D}{\bf k}}{(2\pi)^D}(\partial_{k_b}\epsilon_{\bf k}^{(n)})(\partial_{k_a}\partial_{k_a}f_{0}).
\end{eqnarray} Here, we can use 
\begin{equation}
\frac{\partial f_{0}}{\partial\epsilon_{\bf k}^{(n)}}=\frac{-e^{\left(\epsilon_{\bf k}^{(n)}\!-\!E_{F}\right)/(k_{B}T)}}{\left[1+e^{\left(\epsilon_{\bf k}^{(n)}\!-\!E_{F}\right)/(k_{B}T)}\right]^{2}k_{B}T}=\frac{-1}{2k_{B}T\left\{1+\cosh\left[\left(\epsilon_{\bf k}^{(n)}-E_{F}\right)/(k_{B}T)\right]\right\}}.
\end{equation}
In the above, we have approximated the expressions to the leading order in a small dimensionless parameter $\frac{e\tau}{\hbar}\mathcal{E}_{a}a$ (with the lattice constant $a$) that is proportional to $\tau$.
Taking the Hamiltonian \eqref{eq:H} as an example, the factor $\partial_{k_y}\epsilon_{\bf k}^{(n)}$ in both $C_{yx}$ and $U_{yxx}$ is an odd function of $k_y$, and the factor $\partial_{k_x}f_{0}$ in $C_{yx}$ is an even function of $k_y$. Therefore, the integrand $(\partial_{k_y}\epsilon_{\bf k}^{(n)})(\partial_{k_x} f_{0})$ in $C_{yx}$ is an odd function of $k_y$, which contributes zero with the integral of $k_y$. Similarly, the integrand $(\partial_{k_y}\epsilon_{\bf k}^{(n)})(\partial_{k_x}\partial_{k_x}f_{0})$ in $U_{yxx}$ is an odd function of $k_y$, which also contributes zero with the integral of $k_y$.

\subsection{Expression for the Floquet Hamiltonian~\eqref{eq:H_F}}\label{Methods_2}

In this subsection, we derive the concrete analytical expressions for the time Fourier components ${\cal H}_{0}$, ${\cal H}_{-n}$, and ${\cal H}_{n}$ that enter the Floquet Hamiltonian (\ref{eq:HF0}) of the main text.
With ${\bf A}(t)=\tilde{\omega}^{-1}E_{0}(\sin(\tilde{\omega} t),\sin(\tilde{\omega} t + \varphi))$ and $A_{0}=eE_{0}/(\hbar\tilde{\omega})$, we have the photon-dressed effective Hamiltonian as
\begin{eqnarray}
{\cal H}({\bf k},t)&=&{\cal H}\left({\bf k} - \frac{e}{\hbar}{\bf A}(t)\right) \nonumber\\
&=& t_{0}[k_{x}-A_{0}\sin(\tilde{\omega} t)]\sigma_{0} \!+\! v[k_{y}-A_{0}\sin(\tilde{\omega} t + \varphi)]\sigma_{x} \!+\! \eta v[k_{x}-A_{0}\sin(\tilde{\omega} t)]\sigma_{y} \nonumber\\
&&\!+\! \left\{ m-\alpha\left\{[k_{x}-A_{0}\sin(\tilde{\omega} t)]^{2} \!+\! [k_{y}-A_{0}\sin(\tilde{\omega} t + \varphi)]^{2} \right\} \right\}\sigma_{z} \nonumber\\
&=& t_{0}k_{x}\sigma_{0} \!+\! vk_{y}\sigma_{x} \!+\! \eta vk_{x}\sigma_{y} \!+\! m\sigma_{z} \!-\! A_{0}\frac{t_{0}}{2i}\left(e^{i\tilde{\omega} t} - e^{-i\tilde{\omega} t}\right)\sigma_{0} \nonumber\\
&&\!- A_{0}\frac{v}{2i}\left[e^{i(\tilde{\omega} t + \varphi)} - e^{-i(\tilde{\omega} t + \varphi)}\right]\sigma_{x}\!-\! A_{0}\eta\frac{v}{2i}\left(e^{i\tilde{\omega} t} - e^{-i\tilde{\omega} t}\right)\sigma_{y} \nonumber\\
&&\!- \alpha\left\{\left[k_{x}-\frac{A_{0}}{2i}\left(e^{i\tilde{\omega} t} - e^{-i\tilde{\omega} t}\right)\right]^{2} + \left[k_{y}-\frac{A_{0}}{2i}\left(e^{i(\tilde{\omega} t + \varphi)} - e^{-i(\tilde{\omega} t + \varphi)}\right)\right]^{2} \right\} \sigma_{z}.
\end{eqnarray}
We can extract the time Fourier components in ${\cal H}({\bf k},t)$ to give analytical expressions for ${\cal H}_{0}$, ${\cal H}_{-n}$, and ${\cal H}_{n}$ in the Floquet Hamiltonian (\ref{eq:HF0}):
\begin{eqnarray}
{\cal H}_{0}&\!=\!& \frac{1}{T} \int_{0}^{T}{\cal H}({\bf k},t) dt
\!=\! t_{0}k_{x}\sigma_{0} \!+\! vk_{y}\sigma_{x} \!+\! \eta vk_{x}\sigma_{y} \!+\! \left(m-\alpha A_{0}^{2}-\alpha k^2\right)\sigma_{z}, \label{eq:H0}\\
\!{\cal H}_{-1}&\!=\!& \frac{1}{T} \int_{0}^{T}{\cal H}({\bf k},t) e^{-i\tilde{\omega} t}dt
\!=\! i\frac{A_{0}t_{0}}{2}\sigma_{0} \!+\! i\frac{A_{0}v}{2}e^{i\varphi}\sigma_{x}\!+\! \eta i\frac{A_{0}v}{2}\sigma_{y} \!-\! iA_{0}\alpha\left(k_{x}+k_{y}e^{i\varphi}\right)\sigma_{z}, \label{eq:H-1}\\
\!{\cal H}_{1}&\!=\!& \frac{1}{T} \int_{0}^{T}{\cal H}({\bf k},t) e^{i\tilde{\omega} t}dt
\!=\! -i\frac{A_{0}t_{0}}{2}\sigma_{0} \!-\! i\frac{A_{0}v}{2}e^{-i\varphi}\sigma_{x}\!-\! \eta i\frac{A_{0}v}{2}\sigma_{y} \!+\! iA_{0}\alpha\left(k_{x}+k_{y}e^{-i\varphi}\right)\sigma_{z},\label{eq:H1}\\
\!{\cal H}_{-2}&\!=\!& \frac{1}{T} \int_{0}^{T}{\cal H}({\bf k},t) e^{-i2\tilde{\omega} t}dt
\!=\! \frac{1}{4}A_{0}^{2}\alpha(1+e^{2i\varphi})\sigma_{z}, \label{eq:H-2}\\
\!{\cal H}_{2}&\!=\!& \frac{1}{T} \int_{0}^{T}{\cal H}({\bf k},t) e^{i2\tilde{\omega} t}dt
\!=\! \frac{1}{4}A_{0}^{2}\alpha(1+e^{-2i\varphi})\sigma_{z},\label{eq:H2}\\
\!{\cal H}_{-n<-2}&\!=\!& \frac{1}{T} \int_{0}^{T}{\cal H}({\bf k},t) e^{-in\tilde{\omega} t}dt
\!=\! 0, \label{eq:H-n}\\
\!{\cal H}_{n>2}&\!=\!& \frac{1}{T} \int_{0}^{T}{\cal H}({\bf k},t) e^{in\tilde{\omega} t}dt
\!=\! 0.\label{eq:Hn}
\end{eqnarray}

Importantly, only the $n=1$ commutator in the Floquet-expanded effective Hamiltonian evaluates to nonzero:
\begin{eqnarray}
\frac{[{\cal H}_{-1}, {\cal H}_{1}]}{\hbar\tilde{\omega}}
&=&\frac{A_{0}^{2}}{\hbar\tilde{\omega}}\left\{ \eta\frac{v^2}{4}e^{i\varphi}[\sigma_{x}, \sigma_{y}] + \eta\frac{v^2}{4}e^{-i\varphi}[\sigma_{y}, \sigma_{x}] \right.\nonumber\\
&&\left. -\frac{v\alpha}{2}e^{i\varphi}\left(k_{x}+k_{y}e^{-i\varphi}\right)[\sigma_{x}, \sigma_{z}] -\frac{v\alpha}{2}e^{-i\varphi}\left(k_{x}+k_{y}e^{i\varphi}\right)[\sigma_{z}, \sigma_{x}] \right.\nonumber\\
&&\left. -\eta\frac{v\alpha}{2}\left(k_{x}+k_{y}e^{-i\varphi}\right)[\sigma_{y}, \sigma_{z}] - \eta\frac{v\alpha}{2}\left(k_{x}+k_{y}e^{i\varphi}\right)[\sigma_{z}, \sigma_{y}] \right\} \nonumber\\
&=&\frac{A_{0}^{2}}{\hbar\tilde{\omega}}\left\{ \eta\frac{v^2}{4}e^{i\varphi}[\sigma_{x}, \sigma_{y}] + \eta\frac{v^2}{4}e^{-i\varphi}[\sigma_{y}, \sigma_{x}] \right.\nonumber\\
&&\left. -\frac{v\alpha}{2}\left(k_{x}e^{i\varphi}+k_{y}\right)[\sigma_{x}, \sigma_{z}] -\frac{v\alpha}{2}\left(k_{x}e^{-i\varphi}+k_{y}\right)[\sigma_{z}, \sigma_{x}] \right.\nonumber\\
&&\left. -\eta\frac{v\alpha}{2}\left(k_{x}+k_{y}e^{-i\varphi}\right)[\sigma_{y}, \sigma_{z}] - \eta\frac{v\alpha}{2}\left(k_{x}+k_{y}e^{i\varphi}\right)[\sigma_{z}, \sigma_{y}] \right\} \nonumber\\
&=&\frac{A_{0}^{2}}{\hbar\tilde{\omega}}
\left\{i\eta\frac{v^2}{2}(e^{i\varphi}-e^{-i\varphi})\sigma_{z} + iv\alpha(e^{i\varphi}-e^{-i\varphi})k_{x}\sigma_{y} + i\eta v\alpha(e^{i\varphi}-e^{-i\varphi})k_{y}\sigma_{x} \right\} \nonumber\\
&=&-\frac{A_{0}^{2}(\sin\varphi)}{\hbar\tilde{\omega}}
\left( \eta v^2\sigma_{z} + 2v\alpha k_{x}\sigma_{y} + 2\eta v\alpha k_{y}\sigma_{x} \right),\label{eq:H1-1}\\
\frac{[{\cal H}_{-2}, {\cal H}_{2}]}{2\hbar\tilde{\omega}}
&=&\frac{A_{0}^{2}\alpha}{8\hbar\tilde{\omega}}\left\{ (1+e^{2i\varphi})(1+e^{-2i\varphi})[\sigma_{z}, \sigma_{z}] \right\}=0, \label{eq:H2-2}\\
\frac{[{\cal H}_{-n<-2}, {\cal H}_{n>2}]}{n\hbar\tilde{\omega}}
&=&0, \label{eq:Hn-n}
\end{eqnarray} where we have used $[\sigma_{a},\sigma_{b}]=2i\varepsilon_{abc}\sigma_{c}$.

Therefore, the Floquet Hamiltonian can be written as
\begin{equation}
{\cal H}^{(F)}({\bf k})
\!=\!t_{0}k_{x}\sigma_{0} \!+\! vk_{y}\left(1\!-\!\eta\frac{2\alpha A_{0}^{2}\sin\varphi}{\hbar\tilde{\omega}}\right)\sigma_{x} \!+\! \eta vk_{x}\left(1\!-\!\eta\frac{2\alpha A_{0}^{2}\sin\varphi}{\hbar\tilde{\omega}}\right)\sigma_{y} \!+\! \left(m\!-\!\alpha A_{0}^{2}\!-\!\eta\frac{v^{2}A_{0}^{2}\sin\varphi}{\hbar\tilde{\omega}}\!-\!\alpha k^2\right)\sigma_{z}.\label{eq:Appendix_H_F}
\end{equation}

\subsection{Inversion symmetry}\label{Methods_3}

We show that the Floquet Hamiltonian \eqref{eq:Appendix_H_F} (or Eq.~\eqref{eq:H_F} in the main text) is inversion-symmetric only when $t_0=0$. As such, a nonzero $t_0$ is necessary for the BCD.

The Floquet Hamiltonian~\eqref{eq:Appendix_H_F} under inversion transformation becomes
\begin{eqnarray}\label{eq:inversion}
&&{\cal I}{\cal H}^{(F)}({\bf k}){\cal I}^{-1} \nonumber\\
&\!=\!&t_{0}k_{x}\sigma_{0} \!-\! vk_{y}\left(1\!-\!\eta\frac{2\alpha A_{0}^{2}\sin\varphi}{\hbar\tilde{\omega}}\right)\sigma_{x} \!-\! \eta vk_{x}\left(1\!-\!\frac{2\alpha A_{0}^{2}\sin\varphi}{\hbar\tilde{\omega}}\right)\sigma_{y} \!+\! \left(m\!-\!\alpha A_{0}^{2}\!-\!\eta\frac{v^{2}A_{0}^{2}\sin\varphi}{\hbar\tilde{\omega}}\!-\!\alpha k^2\right)\sigma_{z} \nonumber\\
&\!=\!& {\cal H}^{(F)}(-{\bf k}) \!+\! 2t_{0}k_{x}\sigma_{0},
\end{eqnarray}
where ${\cal I}=\sigma_{z}$~\cite{chen2015magnetoinfrared} is the inversion operator, and we have used
\begin{eqnarray}
&&(\sigma_{z})(\sigma_{0})(\sigma_{z})^{-1}=\sigma_{0},\\
&&(\sigma_{z})(\sigma_{x})(\sigma_{z})^{-1}=-\sigma_{x} ,\\
&&(\sigma_{z})(\sigma_{y})(\sigma_{z})^{-1}=-\sigma_{y} ,\\
&&(\sigma_{z})(\sigma_{z} )(\sigma_{z})^{-1}=\sigma_{z}.
\end{eqnarray}
As a result of Eq.~\eqref{eq:inversion}, if $t_{0}=0$, we have ${\cal I}{\cal H}^{(F)}({\bf k}){\cal I}^{-1}={\cal H}^{(F)}(-{\bf k})$ which satisfies inversion symmetry.
However, for $t_{0} \neq 0$, Eq.~\eqref{eq:inversion}, the extra $2t_0k_x\sigma_0$ term appears due to broken inversion symmetry.

\subsection{Time-reversal symmetry is broken}\label{Methods_4}

Here we show that the Floquet Hamiltonian \eqref{eq:Appendix_H_F} (or Eq.~\eqref{eq:H_F} in the main text) does not satisfy the time-reversal symmetry regardless of the values of $t_0$ and $A_0$.

The Floquet Hamiltonian~\eqref{eq:Appendix_H_F} under time-reversal transformation becomes
\begin{eqnarray}\label{eq:time_reversal}
&&{\cal T}{\cal H}^{(F)}({\bf k}){\cal T}^{-1} \nonumber\\
&\!=\!&t_{0}k_{x}\sigma_{0} \!-\! vk_{y}\left(1\!-\!\eta\frac{2\alpha A_{0}^{2}\sin\varphi}{\hbar\tilde{\omega}}\right)\sigma_{x} \!-\! \eta vk_{x}\left(1\!-\!\frac{2\alpha A_{0}^{2}\sin\varphi}{\hbar\tilde{\omega}}\right)\sigma_{y} \!-\! \left(m\!-\!\alpha A_{0}^{2}\!-\!\eta\frac{v^{2}A_{0}^{2}\sin\varphi}{\hbar\tilde{\omega}}\!-\!\alpha k^2\right)\sigma_{z} \nonumber\\
&\!=\!&{\cal H}^{(F)}(-{\bf k}) \!+\! 2t_{0}k_{x}\sigma_{0} \!-\! 2\left(m\!-\!\alpha A_{0}^{2}\!-\!\eta\frac{v^{2}A_{0}^{2}\sin\varphi}{\hbar\tilde{\omega}}\!-\!\alpha k^2\right)\sigma_{z},
\end{eqnarray}
where ${\cal T}=i\sigma_{y}{\cal K}$~\cite{chen2015magnetoinfrared} is the time-reversal operator with the complex conjugate operator ${\cal K}$ such that ${\cal K}{\cal H}^{(F)}({\bf k}){\cal K}^{-1}={\cal H}^{(F)^*}({\bf k})$, and
\begin{eqnarray}
&&(i\sigma_{y})(\sigma_{0})(i\sigma_{y})^{-1}=\sigma_{0},\\
&&(i\sigma_{y})(\sigma_{x})(i\sigma_{y})^{-1}=-\sigma_{x} ,\\
&&(i\sigma_{y})(K\sigma_{y}K^{-1})(i\sigma_{y})^{-1}=-\sigma_{y} ,\\
&&(i\sigma_{y})(\sigma_{z} )(i\sigma_{y})^{-1}=-\sigma_{z}.
\end{eqnarray}
Due to the presence of the $\sigma_0$ and $\sigma_z$ terms, Eq.~\eqref{eq:time_reversal} shows that the time-reversal symmetry is broken, i.e., ${\cal T}{\cal H}^{(F)}({\bf k}){\cal T}^{-1} \neq {\cal H}^{(F)}(-{\bf k})$ whether $t_{0}$ and $A_{0}$ equal zero or not.
In the same way, we can show that the original static tilted Dirac Hamiltonian [Eq.~\eqref{eq:H} in the main text] also does not satisfy the time-reversal symmetry, regardless of whether $t_{0}$ vanishes.

\subsection{Competition between the linear and nonlinear current densities}\label{Methods_5}

In this subsection, we present more data on the competition between the linear and nonlinear current densities as a function of the Fermi energy at different light intensities.

\begin{figure}[htpb]
\centering
\includegraphics[width=0.45\columnwidth]{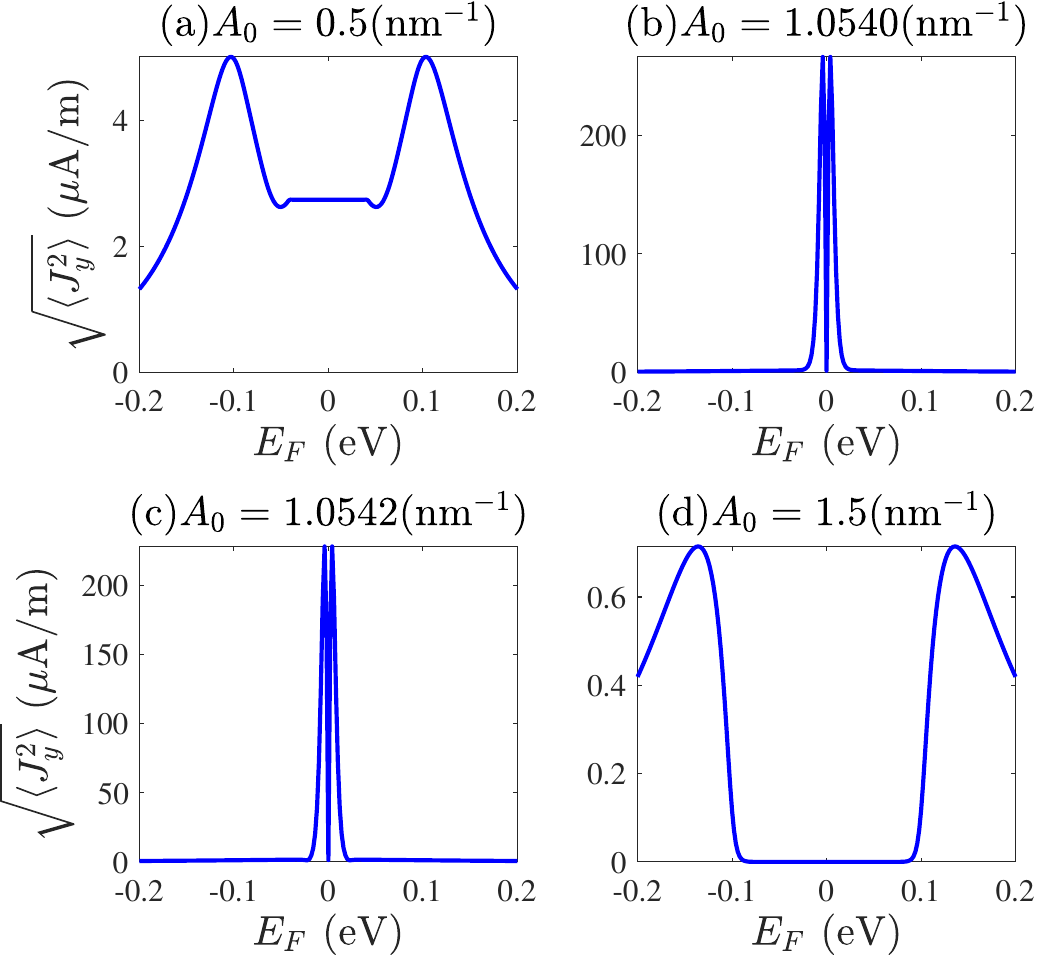}
\caption{\textbf{Competition between the linear and nonlinear current densities.} The root-mean-squrare total current density [Eq.~\eqref{eq:J_total}] along the $y$ direction, averaged over an oscillation period as a function of the Fermi energy $E_F$ at different light intensities from (a) in the Chern insulator phase, to (b,c) just before and after the topological phase transition, and (d) in the trivial phase. Very sharp and divergently large current density peaks occur very close to the phase transition. Other parameters are the electronic field intensity $\mathcal{E}_{x}$=0.1 V/m, $\varphi=\pi/2$ (right-handed circularly polarized light), $\hbar\tilde{\omega}=1$ eV, $t_{0}=0.05$ eV$\cdot$nm, $v=0.1$ eV$\cdot$nm, $\alpha=0.1$ eV$\cdot$nm$^2$, $m=0.1$ eV, $\eta=-1$, $k_{B}T=0.003$ eV, i.e., $T\approx34.8136$ K, $\tau\approx4.12434\times10^{-14}$ s~\cite{ma2019observation}, and $\omega=17.777$ Hz~\cite{ma2019observation}.}\label{fig:Jy_Ef_together}
\end{figure}

As shown in Fig.~\ref{fig:Jy_Ef_together}, the root-mean-square total current density [Eq.~\eqref{eq:J_total} as defined in the main text] along the $y$ direction is plotted as a function of the Fermi energy $E_F$ at different light intensities, for fixed electronic field intensity or amplitude $\mathcal{E}_{x}$=0.1 V/m.

When $A_0$ is sufficiently small such that the system is in the Chern insulator phase, i.e., $A_0=0.5$ nm$^{-1}$, there is an obvious nonzero flat $\sqrt{\langle J_{y}^{2}\rangle}$ region as shown in Fig.~\ref{fig:Jy_Ef_together}(a), which is approximately at the value of the quantized linear Hall current density. The two broad peaks at larger $E_F$ (which are in the bulk bands) are not from the nonlinear Hall response, but rather result from the non-quantized Hall response from incomplete band filling.

But when $A_0$ tends to the critical value $A_{0c}\approx1.0541$ nm$^{-1}$, the gap and hence the flat region disappear, with two very pronounced sharp peaks near the point $E_F=0$ as shown in Figs.~\ref{fig:Jy_Ef_together}(b) and \ref{fig:Jy_Ef_together}(c). While they are also within the bulk bands, teetering at their edges, the peak values of the root-mean-square current density are far higher. That is due mostly to the nonlinear Hall contribution from divergently large BCD. We emphasize that although such a large nonlinear Hall response seemingly requires fine-tuning to observe, the light amplitude $A_0$ is exactly such a very tunable parameter. When $A_0$ becomes even larger such that the system is in the topologically trivial phase with vanishing Chern number, i.e., $A_{0}=1.5$ nm$^{-1}$, there is a flat region at zero as shown in Fig.~\ref{fig:Jy_Ef_together}(d), where both the linear and nonlinear Hall effects essentially vanish.

%%%%%%%%%%%%%%%%%%%%%%%%%%%%%%%%%%%%%%%%%%%%%%%%%%
%%%%%%%%%%%%%%%%%%%%%%%%%%%%%%%%%%%%%%%%%%%%%%%%%%

\end{sloppypar}
\section*{References}
\bibliography{references_nonlinear_TB}

\setcounter{equation}{0}
\setcounter{figure}{0}
\setcounter{table}{0}
\setcounter{section}{0}
\setcounter{secnumdepth}{3}
\renewcommand{\theequation}{S\arabic{equation}}
\renewcommand{\thefigure}{S\arabic{figure}}
\renewcommand{\thesection}{S\Roman{section}}
\renewcommand*{\bibnumfmt}[1]{[S#1]}
\onecolumngrid
\flushbottom
\begin{sloppypar}
\newpage
\section{Supplementary Materials for ``Light-enhanced nonlinear Hall effect"}

\subsection{Supplementary Note 1: Tight-binding model}\label{Supplement_1}

In this subsection, we derive the lattice tight-binding model for the Floquet Hamiltonian \eqref{eq:Appendix_H_F}.
The Floquet Hamiltonian is written as
\begin{equation}
{\cal H}^{(F)}({\bf k})=t_{0}k_x\sigma_{0}+\sum_{i=x,y,z}h_{i}^{(F)}\sigma_{i},\label{eq:Appendix_H_F1}
\end{equation} where
\begin{eqnarray}
h_{x}^{(F)}&\!=&vk_{y}\left(1-\eta\frac{2\alpha A_{0}^{2}\sin\varphi}{\hbar\tilde{\omega}}\right), \\
h_{y}^{(F)}&\!=&\eta vk_{x}\left(1-\eta\frac{2\alpha A_{0}^{2}\sin\varphi}{\hbar\tilde{\omega}}\right), \\
h_{z}^{(F)}&\!=&m-\alpha A_{0}^{2}-\eta\frac{v^{2}A_{0}^{2}\sin\varphi}{\hbar\tilde{\omega}}-\alpha k^2
=m-A_{0}^{2}\left(\alpha + \eta\frac{v^{2}\sin\varphi}{\hbar\tilde{\omega}}\right)-\alpha k^2.
\end{eqnarray}
Here $A_{0}=eE_{0}/(\hbar\tilde{\omega})$ is the electromagnetic potential amplitude.

To regularize the model onto a lattice, one makes the following replacements~\cite{shen2017topological}:
\begin{eqnarray}
&&k_{j}\rightarrow\frac{1}{a_{j}}\sin(k_{j}a_{j}),\label{eq:sin}\\
&&k_{j}^{2}\rightarrow\frac{2}{a_{j}^2}[1-\cos(k_{j}a_{j})],\label{eq:cos}
\end{eqnarray} where $j=x,y$ and $a_j$ is the lattice constant along the $j$ direction.
With the mappings \eqref{eq:sin} and \eqref{eq:cos} and $a_{x}=a_{y}=a$, one obtains
\begin{eqnarray}
h_{0}&\rightarrow&\frac{t_{0}}{a}\sin(k_{x}a),\\
h_{x}^{(F)}&\rightarrow&\frac{v}{a}\sin(k_{y}a)\left(1-\eta\frac{2\alpha A_{0}^{2}\sin\varphi}{\hbar\tilde{\omega}}\right), \\
h_{y}^{(F)}&\rightarrow&\eta \frac{v}{a}\sin(k_{x}a)\left(1-\eta\frac{2\alpha A_{0}^{2}\sin\varphi}{\hbar\tilde{\omega}}\right), \\
h_{z}^{(F)}&\rightarrow& m-A_{0}^{2}\left(\alpha + \eta\frac{v^{2}\sin\varphi}{\hbar\tilde{\omega}}\right)-\frac{2\alpha}{a^2}[2-\cos(k_{x}a)-\cos(k_{y}a)].
\end{eqnarray}
This gives the lattice eigenenergies for the upper ($+$) and lower ($-$) bands as follows:
\begin{equation}
\epsilon_{\mathbf{k}}^{(\pm)}\!=\!h_{0} \!\pm\! \sqrt{[h_{x}^{(F)}]^{2}\!+\![h_{y}^{(F)}]^{2}\!+\![h_{z}^{(F)}]^{2}},\label{eq:Appendix_energy_F}
\end{equation}
which gives Floquet band dispersion velocities according to
\begin{eqnarray}
v_{k_x}^{(\pm)}&=&\frac{1}{\hbar}\frac{\partial\epsilon_{\bf k}^{(\pm)}}{\partial k_x}
=\frac{\partial h_{0}}{\partial k_x}\pm\frac{\left(h_{x}^{(F)}\frac{\partial h_{x}^{(F)}}{\partial k_x} + h_{y}^{(F)}\frac{\partial h_{y}}{\partial k_x} + h_{z}^{(F)}\frac{\partial h_{z}^{(F)}}{\partial k_x} \right)}{\hbar\sqrt{[h_{x}^{(F)}]^{2}\!+\![h_{y}^{(F)}]^{2}\!+\![h_{z}^{(F)}]^{2}}},\\
v_{k_y}^{(\pm)}&=&\frac{1}{\hbar}\frac{\partial\epsilon_{\bf k}^{(\pm)}}{\partial k_y}
=\pm\frac{\left(h_{x}^{(F)}\frac{\partial h_{x}^{(F)}}{\partial k_y} + h_{y}\frac{\partial h_{y}^{(F)}}{\partial k_y} + h_{z}^{(F)}\frac{\partial h_{z}^{(F)}}{\partial k_y} \right)}{\hbar\sqrt{[h_{x}^{(F)}]^{2}\!+\![h_{y}^{(F)}]^{2}\!+\![h_{z}^{(F)}]^{2}}},
\end{eqnarray} where
\begin{eqnarray}
\frac{\partial h_{0}}{\partial k_x}&=&t_{0}\cos(k_{x}a),\\
\frac{\partial h_{x}^{(F)}}{\partial k_x}&=&0,\\
\frac{\partial h_{y}^{(F)}}{\partial k_x}&=&\eta v\cos(k_{x}a)\left(1-\eta\frac{2\alpha A_{0}^{2}\sin\varphi}{\hbar\tilde{\omega}}\right),\\
\frac{\partial h_{z}^{(F)}}{\partial k_x}&=&-\frac{2\alpha}{a}\sin(k_{x}a),\\
\frac{\partial h_{x}^{(F)}}{\partial k_y}&=&v\cos(k_{y}a)\left(1-\eta\frac{2\alpha A_{0}^{2}\sin\varphi}{\hbar\tilde{\omega}}\right),\\
\frac{\partial h_{y}^{(F)}}{\partial k_y}&=&0,\\
\frac{\partial h_{z}^{(F)}}{\partial k_y}&=&-\frac{2\alpha}{a}\sin(k_{y}a).
\end{eqnarray}

\subsection{Supplementary Note 2: Berry curvature}\label{Supplement_2}

For pedagogical purposes, we shall derive the analytical expression for the Berry curvature of our Floquet lattice tight-binding model Hamiltonian~\eqref{eq:Appendix_H_F1} through two different methods. Both are generically applicable to any two-band model, although the computational complexities differ.

\subsubsection{Method one}\label{Supplement_2_1}

We first show that the traceless part of any two-band Hamiltonian, such as our Hamiltonian in Eq.~\eqref{eq:Appendix_H_F1},
can be parametrized in terms of quantities $R$ and $D$. Explicitly, we start by decomposing the model as ${\cal H}^{(F)}=h_{0}\sigma_{0}+h_{+}^{(F)}\sigma_{+} + h_{-}^{(F)}\sigma_{-} + h_{z}^{(F)}\sigma_z=h_{0}\sigma_{0}+[(h_{+}^{(F)} + h_{-}^{(F)})/2]\sigma_{x} + [i(h_{+}^{(F)} - h_{-}^{(F)})/2]\sigma_{y} + h_{z}^{(F)}\sigma_{z}$ with $h_{\pm}^{(F)}=h_{x}^{(F)} \mp ih_{y}^{(F)}$ and $\sigma_{\pm}=(\sigma_{x} \pm i\sigma_{y})/2$, such that we can perform the following parametrization on the Bloch sphere~\cite{qin2024kinked}:
\begin{eqnarray}
h_x^{(F)}&=&h^{(F)}\sin\theta\cos\phi=h^{(F)}\sqrt{1 - D^2}\cos\phi,\\
h_y^{(F)}&=&h^{(F)}\sin\theta\sin\phi=h^{(F)}\sqrt{1 - D^2}\sin\phi,\\
h_z^{(F)}&=&h^{(F)}\cos\theta=h^{(F)}D,\\
h^{(F)}&=&\sqrt{[h_x^{(F)}]^2 + [h_y^{(F)}]^2 + [h_z^{(F)}]^2},\\
D&=&\frac{h_z^{(F)}}{\sqrt{[h_x^{(F)}]^2 + [h_y^{(F)}]^2 + [h_z^{(F)}]^2}},\\
R&=&\sqrt{\frac{h_{+}^{(F)}}{h_{-}^{(F)}}}=\sqrt{\frac{h_x^{(F)} - ih_y^{(F)}}{h_x^{(F)} + ih_y^{(F)}}}=\sqrt{\frac{e^{-i\phi}}{e^{i\phi}}}=e^{-i\phi},
\end{eqnarray} where we use $h_x^{(F)} + ih_y^{(F)}=h_{\perp}^{(F)}(\cos\phi + i\sin\phi)=h_{\perp}^{(F)}e^{i\phi}$ and $h_x^{(F)} - ih_y^{(F)}=h_{\perp}^{(F)}(\cos\phi - i\sin\phi)=h_{\perp}^{(F)}e^{-i\phi}$ with $h_{\perp}^{(F)}=\sqrt{[h_x^{(F)}]^2+[h_y^{(F)}]^2}$.
Hence, our Hamiltonian can be expressed as
\begin{eqnarray}
{\cal H}^{(F)}&=&h_{0}\sigma_{0} + h_x^{(F)}\sigma_x + h_y^{(F)}\sigma_y + h_z^{(F)}\sigma_z
=h_{0}\sigma_{0} + \begin{pmatrix}
h_z^{(F)} & h_x^{(F)} - ih_y^{(F)}  \\
h_x^{(F)} + ih_y^{(F)} & - h_z^{(F)}
\end{pmatrix} \nonumber\\
&=&h_{0}\sigma_{0} + \begin{pmatrix}
h^{(F)}D & h^{(F)}\sqrt{1 - D^2}(\cos\phi - i\sin\phi) \\
h^{(F)}\sqrt{1 - D^2}(\cos\phi + i\sin\phi) & - h^{(F)}D
\end{pmatrix} \nonumber\\
&=&h_{0}\sigma_{0} + h^{(F)}\begin{pmatrix}
D & \sqrt{1 - D^2}e^{-i\phi} \\
\sqrt{1 - D^2}e^{i\phi} & - D
\end{pmatrix}
=h_{0}\sigma_{0} + h^{(F)}\begin{pmatrix}
D & \sqrt{1 - D^2}R \\
\sqrt{1 - D^2}R^{-1} & - D
\end{pmatrix}.
\end{eqnarray}

Explicitly, its right and left eigenstates are expressible entirely in terms of $R$ and $D$:
\begin{eqnarray}
|\psi^{(\pm)}\rangle&=&\frac{1}{\sqrt{2}}\begin{pmatrix}
\pm R\sqrt{1\pm D} \\
\sqrt{1\mp D}
\end{pmatrix}, \\
\langle\psi^{(\pm)}|&=&\frac{1}{\sqrt{2}}\begin{pmatrix}
\pm \sqrt{1\pm D}/R &
\sqrt{1\mp D}
\end{pmatrix}.
\end{eqnarray}
From this, we obtain the Berry curvature as
\begin{eqnarray}
\Omega_{{\bf k},z}^{(+)}&=&i\left( \langle\partial_{k_x}\psi^{(+)}|\partial_{k_y}\psi^{(+)}\rangle - \langle\partial_{k_y}\psi^{(+)}|\partial_{k_x}\psi^{(+)}\rangle \right) \nonumber\\
&=&\frac{i}{2}\left[ \begin{pmatrix}
\partial_{k_x}(\sqrt{1+ D}/R) &
\partial_{k_x}\sqrt{1- D}
\end{pmatrix}\begin{pmatrix}
\partial_{k_y}( R\sqrt{1+ D}) \\
\partial_{k_y}\sqrt{1- D}
\end{pmatrix}  - \begin{pmatrix}
\partial_{k_y}(\sqrt{1+ D}/R) &
\partial_{k_y}\sqrt{1- D}
\end{pmatrix}\begin{pmatrix}
\partial_{k_x}( R\sqrt{1+ D}) \\
\partial_{k_x}\sqrt{1- D}
\end{pmatrix}  \right] \nonumber\\
&=&\frac{i}{2}\left[
\partial_{k_x}(\sqrt{1\!+\! D}/R)\partial_{k_y}( R\sqrt{1\!+\!D}) \!+\!
\partial_{k_x}\sqrt{1\!-\! D}\partial_{k_y}\sqrt{1\!-\! D} \!-\!
\partial_{k_y}(\sqrt{1\!+\! D}/R)\partial_{k_x}( R\sqrt{1\!+\!D}) \!-\!
\partial_{k_y}\sqrt{1\!-\! D}\partial_{k_x}\sqrt{1\!-\! D} \right] \nonumber\\
&=&\frac{i}{2}\left[
\partial_{k_x}(\sqrt{1\!+\! D}/R)\partial_{k_y}( R\sqrt{1\!+\!D}) \!-\!
\partial_{k_y}(\sqrt{1\!+\! D}/R)\partial_{k_x}( R\sqrt{1\!+\!D}) \right] \nonumber\\
&=&\frac{i}{2}\left[
\left(\frac{1}{2R\sqrt{1\!+\! D}}\partial_{k_x}D - \frac{\sqrt{1\!+\! D}}{R^2}\partial_{k_x}R \right)\left( \frac{R}{2\sqrt{1\!+\!D}}\partial_{k_y}D + \sqrt{1\!+\!D}\partial_{k_y}R \right) \right.\nonumber\\
&&\left. \!-\!
\left(\frac{1}{2R\sqrt{1\!+\! D}}\partial_{k_y}D - \frac{\sqrt{1\!+\! D}}{R^2}\partial_{k_y}R \right)\left( \frac{R}{2\sqrt{1\!+\!D}}\partial_{k_x}D + \sqrt{1\!+\!D}\partial_{k_x}R \right) \right] \nonumber\\
&=&\frac{i}{2R}\left[(\partial_{k_x}D)(\partial_{k_y}R)  -  (\partial_{k_y}D)(\partial_{k_x}R) \right]
=\frac{i}{2}\left[(\partial_{k_x}D)(\partial_{k_y}\ln R)  -  (\partial_{k_y}D)(\partial_{k_x}\ln R) \right],
\end{eqnarray}
\begin{eqnarray}
\Omega_{{\bf k},z}^{(-)}&=&i\left( \langle\partial_{k_x}\psi^{(-)}|\partial_{k_y}\psi^{(-)}\rangle - \langle\partial_{k_y}\psi^{(-)}|\partial_{k_x}\psi^{(-)}\rangle \right) \nonumber\\
&=& \frac{i}{2}\left[ \begin{pmatrix}
\partial_{k_x}(-\sqrt{1- D}/R) &
\partial_{k_x}\sqrt{1+ D}
\end{pmatrix}\begin{pmatrix}
\partial_{k_y}( -R\sqrt{1- D}) \\
\partial_{k_y}\sqrt{1+ D}
\end{pmatrix}  - \begin{pmatrix}
\partial_{k_y}(-\sqrt{1- D}/R) &
\partial_{k_y}\sqrt{1+ D}
\end{pmatrix}\begin{pmatrix}
\partial_{k_x}( -R\sqrt{1- D}) \\
\partial_{k_x}\sqrt{1+ D}
\end{pmatrix}  \right] \nonumber\\
&=&\frac{i}{2}\left[
\partial_{k_x}(\sqrt{1\!-\! D}/R)\partial_{k_y}( R\sqrt{1\!-\!D}) \!+\!
\partial_{k_x}\sqrt{1\!+\! D}\partial_{k_y}\sqrt{1\!+\! D} \!-\!
\partial_{k_y}(\sqrt{1\!-\! D}/R)\partial_{k_x}( R\sqrt{1\!-\!D}) \!-\!
\partial_{k_y}\sqrt{1\!+\! D}\partial_{k_x}\sqrt{1\!+\! D} \right] \nonumber\\
&=&\frac{i}{2}\left[
\partial_{k_x}(\sqrt{1\!-\! D}/R)\partial_{k_y}( R\sqrt{1\!-\!D}) \!-\!
\partial_{k_y}(\sqrt{1\!-\! D}/R)\partial_{k_x}( R\sqrt{1\!-\!D}) \right] \nonumber\\
&=&\frac{i}{2}\left[
\left(\frac{1}{2R\sqrt{1\!-\! D}}\partial_{k_x}(-D) - \frac{\sqrt{1\!-\! D}}{R^2}\partial_{k_x}R \right)\left( \frac{R}{2\sqrt{1\!-\!D}}\partial_{k_y}(-D) + \sqrt{1\!-\!D}\partial_{k_y}R \right) \right.\nonumber\\
&&\left. \!-\!
\left(\frac{1}{2R\sqrt{1\!-\! D}}\partial_{k_y}(-D) - \frac{\sqrt{1\!-\! D}}{R^2}\partial_{k_y}R \right)\left( \frac{R}{2\sqrt{1\!-\!D}}\partial_{k_x}(-D) + \sqrt{1\!-\!D}\partial_{k_x}R \right) \right] \nonumber\\
&=&-\frac{i}{2R}\left[(\partial_{k_x}D)(\partial_{k_y}R)  -  (\partial_{k_y}D)(\partial_{k_x}R) \right]
=-\frac{i}{2}\left[(\partial_{k_x}D)(\partial_{k_y}\ln R)  -  (\partial_{k_y}D)(\partial_{k_x}\ln R) \right].
\end{eqnarray}
Finally, substituting in our specific model~\eqref{eq:Appendix_H_F1}, the analytical expression for the Berry curvature is
\begin{eqnarray}
\Omega_{{\bf k},z}^{(\pm)}
&=&\pm\frac{i}{2}\left[(\partial_{k_x}D)(\partial_{k_y}\ln R)  -  (\partial_{k_y}D)(\partial_{k_x}\ln R) \right] \nonumber\\
&=&\pm\frac{\eta \tilde{v}^{2} \left\{\left( \tilde{m} - \frac{4\alpha}{a^2} \right)\cos(k_{x}a)\cos(k_{y}a) + \frac{2\alpha}{a^2}\left[\cos(k_{x}a)+\cos(k_{y}a)\right] \right\}}{2 \left\{\tilde{M}^{2}({\bf k})+\frac{\tilde{v}^{2}}{a^{2}}\left[\sin^{2}(k_{x}a)+\sin^{2}(k_{y}a)\right]\right\}^{1/2} } \nonumber\\
&&\times\frac{1}{\left\{ \left( \tilde{m} - \frac{4\alpha}{a^2} \right)^{2} + \left(\frac{2\alpha}{a^2}\right)^{2}\left[\cos(k_{x}a)+\cos(k_{y}a)\right]^{2} + \frac{4\alpha}{a^2}\left( \tilde{m} - \frac{4\alpha}{a^2} \right)\left[\cos(k_{x}a)+\cos(k_{y}a)\right] + \frac{\tilde{v}^{2}}{a^{2}}\left[\sin^{2}(k_{x}a)+\sin^{2}(k_{y}a)\right] \right\}} \nonumber\\
&=&\pm\frac{\eta \tilde{v}^{2} \left\{\left( \tilde{m} - \frac{4\alpha}{a^2} \right)\cos(k_{x}a)\cos(k_{y}a) + \frac{2\alpha}{a^2}\left[\cos(k_{x}a)+\cos(k_{y}a)\right] \right\}}{2 \left\{\tilde{M}^{2}({\bf k})+\frac{\tilde{v}^{2}}{a^{2}}\left[\sin^{2}(k_{x}a)+\sin^{2}(k_{y}a)\right]\right\}^{1/2} } \nonumber\\
&&\times\frac{1}{\left\{ \left[ \tilde{m} - \frac{4\alpha}{a^2} + \frac{2\alpha}{a^2}\left[\cos(k_{x}a)+\cos(k_{y}a)\right] \right]^{2} + \frac{\tilde{v}^{2}}{a^{2}}\left[\sin^{2}(k_{x}a)+\sin^{2}(k_{y}a)\right] \right\}} \nonumber\\
&=&\pm\frac{\eta \tilde{v}^{2} \left\{\left( \tilde{m} - \frac{4\alpha}{a^2} \right)\cos(k_{x}a)\cos(k_{y}a) + \frac{2\alpha}{a^2}\left[\cos(k_{x}a)+\cos(k_{y}a)\right] \right\}}{2 \left\{\tilde{M}^{2}({\bf k})+\frac{\tilde{v}^{2}}{a^{2}}\left[\sin^{2}(k_{x}a)+\sin^{2}(k_{y}a)\right]\right\}^{3/2} },
\end{eqnarray} where
\begin{eqnarray}
\tilde{m}&=&m-A_{0}^{2}\left(\alpha + \eta\frac{v^{2}\sin\varphi}{\hbar\tilde{\omega}}\right),\\
\tilde{M}({\bf k})&=&\tilde{m}-\frac{2\alpha}{a^2}\left[2 - \cos(k_{x}a) - \cos(k_{y}a) \right],\\
\tilde{v}&=&v\left(1-\eta\frac{2\alpha A_{0}^{2}\sin\varphi}{\hbar\tilde{\omega}}\right).
\end{eqnarray}

\subsubsection{Method two}\label{Supplement_2_2}

From the expression of the Chern number
\begin{equation}\label{MTChern}
\mathcal{C}=\frac{1}{4\pi}\int_{\text{BZ}}d^{2}\mathbf{k}~\frac{\mathbf{h}^{(F)}}{|\mathbf{h}^{(F)}|^{3}}\cdot(\partial_{k_{x}}\mathbf{h}^{(F)}\times\partial_{k_{y}}\mathbf{h}^{(F)}),
\end{equation}
where BZ stands for the Brillouin zone, we can compute the Berry curvature directly in terms of ${\bold h}^{(F)}$. We have
\begin{eqnarray}
&&\partial_{k_{x}}\mathbf{h}^{(F)}\times\partial_{k_{y}}\mathbf{h}^{(F)} \nonumber\\
&\!=\!&\varepsilon_{ijk}\partial_{k_{x}}h_{j}^{(F)}\partial_{k_{y}}h_{k}^{(F)}=\left|\begin{array}{ccc}
 \mathbf{e}_{x} & \mathbf{e}_{y} & \mathbf{e}_{z} \\
 \partial_{k_{x}}h_{x}^{(F)} & \partial_{k_{x}}h_{y}^{(F)} & \partial_{k_{x}}h_{z}^{(F)} \\
 \partial_{k_{y}}h_{x}^{(F)} & \partial_{k_{y}}h_{y}^{(F)} & \partial_{k_{y}}h_{z}^{(F)}
\end{array}\right|\nonumber\\
&\!=\!&(\partial_{k_{x}}h_{y}^{(F)}\partial_{k_{y}}h_{z}^{(F)}-\partial_{k_{x}}h_{z}^{(F)}\partial_{k_{y}}h_{y}^{(F)})\mathbf{e}_{x}
+(\partial_{k_{x}}h_{z}^{(F)}\partial_{k_{y}}h_{x}^{(F)}-\partial_{k_{x}}h_{x}^{(F)}\partial_{k_{y}}h_{z}^{(F)})\mathbf{e}_{y} \nonumber\\
&&+(\partial_{k_{x}}h_{x}^{(F)}\partial_{k_{y}}h_{y}^{(F)}-\partial_{k_{x}}h_{y}^{(F)}\partial_{k_{y}}h_{x}^{(F)})\mathbf{e}_{z}\nonumber\\
&\!=\!&[(\partial_{k_{x}}h_{y}^{(F)}\partial_{k_{y}}h_{z}^{(F)}\!-\!\partial_{k_{x}}h_{z}^{(F)}\partial_{k_{y}}h_{y}^{(F)}),
(\partial_{k_{x}}h_{z}^{(F)}\partial_{k_{y}}h_{x}^{(F)}\!-\!\partial_{k_{x}}h_{x}^{(F)}\partial_{k_{y}}h_{z}^{(F)}),
(\partial_{k_{x}}h_{x}^{(F)}\partial_{k_{y}}h_{y}^{(F)}\!-\!\partial_{k_{x}}h_{y}^{(F)}\partial_{k_{y}}h_{x}^{(F)})],\nonumber\\
\end{eqnarray}
\begin{eqnarray}
&&\mathbf{h}^{(F)}\cdot(\partial_{k_{x}}\mathbf{h}^{(F)}\times\partial_{k_{y}}\mathbf{h}^{(F)}) \nonumber\\
&=&h_{i}^{(F)}\varepsilon_{ijk}\partial_{k_{x}}h_{j}^{(F)}\partial_{k_{y}}h_{k}^{(F)}\nonumber\\
&=&h_{x}^{(F)}(\partial_{k_{x}}h_{y}^{(F)}\partial_{k_{y}}h_{z}^{(F)}-\partial_{k_{x}}h_{z}^{(F)}\partial_{k_{y}}h_{y}^{(F)})+
h_{y}^{(F)}(\partial_{k_{x}}h_{z}^{(F)}\partial_{k_{y}}h_{x}^{(F)}-\partial_{k_{x}}h_{x}^{(F)}\partial_{k_{y}}h_{z}^{(F)})\nonumber\\
&&+
h_{z}^{(F)}(\partial_{k_{x}}h_{x}^{(F)}\partial_{k_{y}}h_{y}^{(F)}-\partial_{k_{x}}h_{y}^{(F)}\partial_{k_{y}^{(F)}}h_{x}^{(F)}).
\end{eqnarray}
Such that the Berry curvature for our specific model~\eqref{eq:Appendix_H_F1} is given by~\cite{ning2023robustness}
\begin{eqnarray}
\Omega_{{\bf k},z}^{(\pm)}
&=&\mp\frac{1}{2}\frac{\mathbf{h}^{(F)}}{|\mathbf{h}^{(F)}|^{3}}\cdot(\partial_{k_{x}}\mathbf{h}^{(F)}\times\partial_{k_{y}}\mathbf{h}^{(F)}) \nonumber\\
&=&\pm\frac{\eta \tilde{v}^{2} \left\{\left( \tilde{m} - \frac{4\alpha}{a^2} \right)\cos(k_{x}a)\cos(k_{y}a) + \frac{2\alpha}{a^2}\left[\cos(k_{x}a)+\cos(k_{y}a)\right] \right\}}{2 \left\{\tilde{M}^{2}({\bf k})+\frac{\tilde{v}^{2}}{a^{2}}\left[\sin^{2}(k_{x}a)+\sin^{2}(k_{y}a)\right]\right\}^{3/2} }.
\end{eqnarray}

%\newpage
%\clearpage
\subsection{Supplementary Note 3: Energy spectrum, Berry curvature, Hall conductance, and BCD under left-handed circularly polarized light}\label{Supplement_3}

Here, we present the Floquet driving and quench results when left-handed ($\varphi=-\pi/2$) instead of right-handed ($\varphi=\pi/2$) polarized light is used. The results are qualitatively similar to those in the main text, apart from some subtle differences. The main purpose of this subsection is to show that there is negligible difference in the results between the right-handed and the left-handed circularly polarized lights, except for the critical intensity of light $A_{0c}$.
%%%%%%%%%%%%%%%%%%%%%%%%%%%%%%%%%%%%%%%%%%%%%%%%%%
\begin{figure*}[htpb]
\centering
\includegraphics[width=0.8\textwidth]{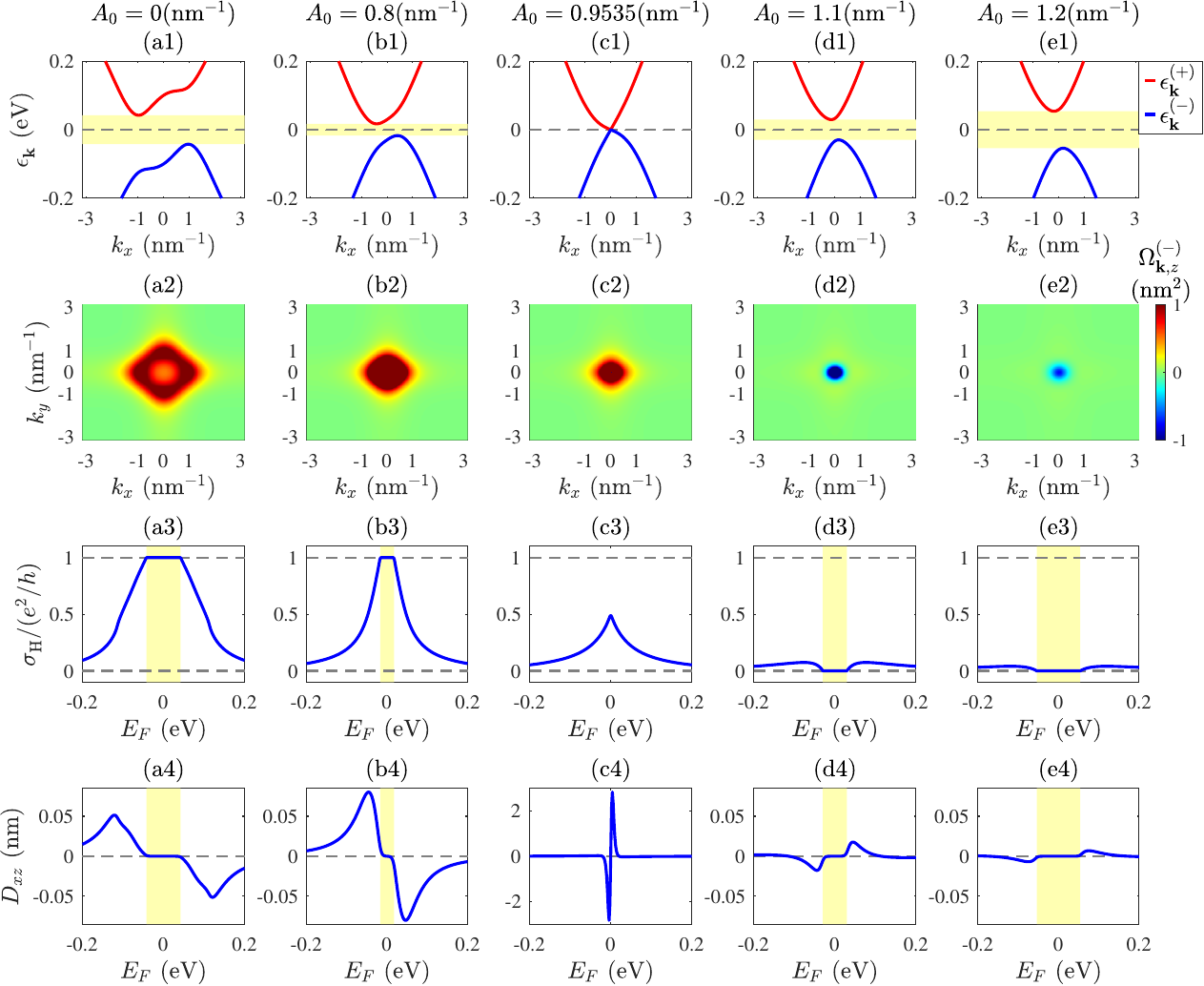}
\caption{\textbf{Floquet band structure [Eq.~\eqref{eq:energy}], Berry curvature [Eq.~\eqref{eq:BC}], Hall conductance [Eq.~\eqref{eq:Hall_0}], and BCD [Eq.~\eqref{eq:BCD_0}] of the nonlinear Hall materials from the Chern insulator phase to the normal insulator phase.} First row (a1)-(e1): Energy spectrum as a function of $k_x$ with $k_y=0$ under different light intensities: (a1) $A_{0}=0$ nm$^{-1}$, (b1) $A_{0}=0.8$ nm$^{-1}$, (c1) $A_{0}\approx A_{0c}\approx0.9535$ nm$^{-1}$, (d1) $A_{0}=1.1$ nm$^{-1}$, and (e1) $A_{0}=1.2$ nm$^{-1}$. Second row (a2)-(e2): Berry curvature of the lower band as a function of $k_x$ and $k_y$ under different light intensities. Third row (a3)-(e3): Hall conductance as a function of the Fermi energy $E_F$ under different light intensities. Fourth row (a4)-(e4): BCD as a function of the Fermi energy $E_F$ under different light intensities. The other parameters are $\varphi=-\pi/2$ (left-handed circularly polarized light), $\hbar\tilde{\omega}=1$ eV, $t_{0}=0.05$ eV$\cdot$nm, $v=0.1$ eV$\cdot$nm, $\alpha=0.1$ eV$\cdot$nm$^2$, $m=0.1$ eV, $\eta=-1$, and $k_{B}T=0.003$ eV, i.e., $T\approx34.8136$ K. These parameters are of the same order as those in two-dimensional massive Dirac models~\cite{ma2019observation,muechler2016topological,liu2021short}.}\label{fig:E_BC_C_BCD_phi_m05pi_together}
\end{figure*}
%%%%%%%%%%%%%%%%%%%%%%%%%%%%%%%%%%%%%%%%%%%%%%%%%%

%%%%%%%%%%%%%%%%%%%%%%%%%%%%%%%%%%%%%%%%%%%%%%%%%%
\begin{figure}[htpb]
\centering
\includegraphics[width=.5\textwidth]{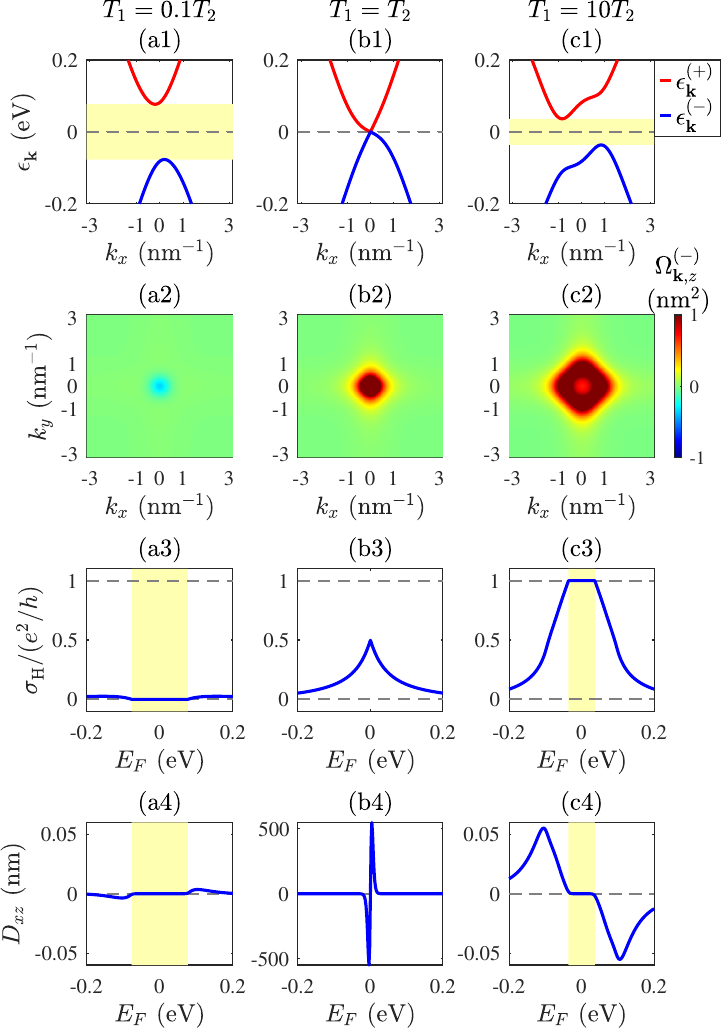}
\caption{\textbf{Floquet quench band structure [Eq.~\eqref{eq:Heff}], Berry curvature [Eq.~\eqref{eq:BC}], Hall conductance [Eq.~\eqref{eq:Hall_0}], and BCD [Eq.~\eqref{eq:BCD_0}] are quenched from the normal insulator phase to the Chern insulator phase.} (a1-c1) Floquet energy bands with $k_{y}=0$. In (a1) $T_{1}=0.1T_{2}$. The ${\cal H}_{2}({\bf k})$ is dominant, gapping the energy band (yellow). In (b1), $T_{1}=T_{2}$. The contributions from ${\cal H}_{1}({\bf k})$ and ${\cal H}_{2}({\bf k})$ cancel each other out, so that the energy band is gapless. In (c1), $T_{1}=10T_{2}$. The ${\cal H}_{1}({\bf k})$ is dominant, gapping the energy band (yellow). Here ${\cal H}_{1}({\bf k})$ denotes the Hamiltonian without light, i.e., $A_{0}=0$, and ${\cal H}_{2}({\bf k})$ denotes the Hamiltonian under right-handed circularly polarized light with $\varphi=-\pi/2$ and $A_{0}=\sqrt{2}A_{0c}=\sqrt{2m/\left(\alpha + \eta\frac{v^{2}\sin\varphi}{\hbar\tilde{\omega}}\right)}\approx1.3484$ nm$^{-1}$. (a2-c2) Berry curvature of the lower band as a function of $k_x$ and $k_y$. (a3-c3) Hall conductance as a function of the Fermi energy $E_{F}$. (a4-c4) BCD as a function of the Fermi energy $E_{F}$. Here, $T_{2}=0.1\hbar/$eV$\approx6.58212\times10^{-17}$ s and the other parameters are the same as those in Fig.~\ref{fig:quench_E_BC_C_BCD_phi05pi_together}.} \label{fig:quench_E_BC_C_BCD_phi_m05pi_together}
\end{figure}
%%%%%%%%%%%%%%%%%%%%%%%%%%%%%%%%%%%%%%%%%%%%%%%%%%

\end{sloppypar}

\end{document}